\RequirePackage{fix-cm}
\documentclass[smallextended,numbook]{svjour3}
\usepackage[utf8]{inputenc}
\usepackage[T1]{fontenc}
\usepackage{lineno}
\usepackage[hidelinks]{hyperref}
\modulolinenumbers[5]

\usepackage{tikz}
\usepackage{amsmath, amssymb, amsfonts}
\usepackage{mathtools}
\usepackage{graphicx}
\usepackage[export]{adjustbox}
\usepackage[percent]{overpic}
\usepackage{paralist}
\usepackage{textcomp}
\usepackage{booktabs}
\usepackage{xcolor}
\usepackage{flafter}

\usepackage{todonotes}
\usepackage[toc,title,page]{appendix}

\usepackage{cleveref} 
\let\cref\Cref

\DeclareMathOperator*{\argmax}{arg\,max}

\usepackage[url=false, style=nature, natbib=true,citestyle=authoryear]{biblatex}
\let\citet\textcite
\let\cite\citep
\newcommand{\methodname}{Research Topic Flows}
\newcommand{\methodshort}{RTF}
\newcommand{\tfn}{Topic Flow Network}
\newcommand{\tfns}{TFN}
\usepackage{colonequals}
\newcommand*{\logeq}{\ratio\Longleftrightarrow}

\let\phi\varphi

\begin{document}
\title{Research Topic Flows in Co-Authorship Networks}

\author{Bastian Schäfermeier \and
    Johannes Hirth \and 
    Tom Hanika}

\institute{Bastian Schaefermeier, Johannes Hirth, Tom Hanika \at
  Knowledge and Data Engineering Group, University of Kassel, Wilhelmshoeher Allee 73, Kassel, Germany.
              \email{\{schaefermeier, hirth, tom.hanika\}@cs.uni-kassel.de}           
              \and
}

\maketitle
\begin{abstract}
In scientometrics, scientific collaboration is often analyzed by means of co-authorships. An aspect which is often overlooked and more difficult to quantify is the flow of expertise between authors from different research topics, which is an important part of scientific progress. With the \emph{Topic Flow Network} (TFN) we propose a graph structure for the analysis of research topic flows between scientific authors and their respective research fields.

Based on a multi-graph and a topic model, our proposed network structure accounts for intratopic as well as intertopic flows. Our method requires for the construction of a TFN solely a corpus of publications (i.e., author and abstract information). From this, research topics are discovered automatically through non-negative matrix factorization. The thereof derived TFN allows for the application of social network analysis techniques, such as common metrics and community detection. Most importantly, it allows for the analysis of intertopic flows on a large, macroscopic scale, i.e., between research topic, as well as on a microscopic scale, i.e., between certain sets of authors.

We demonstrate the utility of \tfns s by applying our method to two comprehensive corpora of altogether 20 Mio. publications spanning more than 60 years of research in the fields computer science and mathematics. Our results give evidence that \tfn s are suitable, e.g., for the analysis of topical communities, the discovery of important authors in different fields, and, most notably, the analysis of intertopic flows, i.e., the transfer of topical expertise. Besides that, our method opens new directions for future research, such as the investigation of influence relationships between research fields.
\end{abstract}


\section{Introduction}
Scientific collaboration is a key factor for improving publication
quality~\cite{collabquality}, it is increasing in
frequency~\cite{sonnenwald2007scientific} and a necessity for
cross-domain research progress. Networks (or graphs) derived from
co-authorship publication data are the most common approach to
investigate scientific collaborations~\cite{newman2001scientific} and
therefore constitute an essential tool to reveal patterns of
collaboration~\cite{newmannCollab}.

Often these investigations are restricted to particular sets of
research topics, e.g.,~\citet{collabPatternPopGen} investigated
collaboration networks in the field of theoretical population genetics
and~\citet{hou2008structure} analyzed the co-authorship network of
Scientometrics journal authors. 
Yet, more often analyses are rather agnostic to the
research field and do focus on other aspects, such as geographical
features~\cite{katz1994geographical}, political demarcations in the
world, e.g.,~\citet{he2009international}, or observed shock events,
such as geopolitical change on the research
landscape~\cite{MoreInternationalAfter1990}.

An increasingly strongly investigated field of research is the analysis of topics
of publications, their emergence, dynamic as well as
specializations. Thereby, approaches to the topic analysis of
publication corpora differ into intrinsic
methods~\cite{churchill2018temporal,rosvall2010mapping} and
extensional ones, i.e., they separately compute an author
collaboration network and topic model for the related publications
~\cite{AuthortopicsJeongLGC20}. Clearly, the combination of
collaboration networks and topic-based models offers new insights for
scientometric analyses of publication corpora, especially an
author-based measurement of topic flows.  Although implicitly the
properties of co-authorships and research topics have certainly been
blended in studies, no explicit modeling of the two as a combined
analysis structure for \emph{\methodname{}} has yet taken
place. Provided that such a structure can be explained and calculated
(in a mathematically justifiable way), it will allow a variety of new
scientometric analysis approaches:
\begin{inparaenum}
  \item It allows to detect inter-topic collaborations between researchers;
  \item Based on the detection, their frequency and intensity can be measured and tracked through time;
  \item In total, all these measurements can be aggregated and a
    comprehensive model for measuring the flow between research topics
    can be derived.
\end{inparaenum}

In this paper, we propose how this combination can be designed in a
mathematically tractable way. Considering human comprehensibility and
explainability we employ an
interpretable \emph{topic model} (non-negative matrix factorization)
for the construction of the \emph{\tfn{}} (TFN). Our approach overcomes
several obstacles, most importantly
\begin{inparaenum}
  \item TFN reflects the variety of research topics of authors and their publications;
  \item TFN can capture the different thematic flows between authors and,
    in an aggregated form, between topics themselves.
\end{inparaenum}

In detail, we introduce a topic model enhanced graph structure to
 investigate how topical expertise flows through collaboration
 networks (i.e., co-author networks). For this, we start
 with a research corpus from which we derive a topic model, that allows for the creation of a directed,
 edge-weighted multi-graph, the \tfns{}. The predicate for a directed edge from author $a$ to author $b$ is that they collaborated on topic $t$ and the expertise of $a$ on $t$ is higher than the expertise of $b$ on $t$.
Based on this structure we are able to measure the flow between research topics by aggregation and thus indirectly the flow of knowledge between research areas.

We study our method on two comprehensive publication corpora from the research fields mathematics and computer science, which amount to a total of 20 Mio. research articles and about 900,000 authors. Both span a period of more than sixty years, starting from 1960. The thereof constructed \tfns s are used in example analyses, such as calculating common social network metrics, (topical) community detection and the recognition of influential authors via PageRank. Most importantly, we derive topic flows for all topics and years in our data set and visualize and interpret five examples.

An advantage of our approach is that it solely rests on the availability of co-authorship information and paper abstracts, i.e., no citation information is required, which is usually more difficult to acquire. Combined with the inherent interpretability of the topic model, this renders the \tfn{} a versatile and comprehensible research tool in the field of scientometrics.

\section{Related Work}
Our presented work draws mainly from research results from the analysis of social (co-authorship) networks, topics therein and recent topic flow modelings. Work on the former is extensive and we want to recollect therefore only the most relevant results for our work. In contrast, our compilation of topic flow approaches is more comprehensive. 

\paragraph{Co-Authorship  Networks}
Co-authorship is one of the best documented properties in scientometrics. Data
based on this attribute are comparatively easy to obtain for a
plethora of research areas, unlike data based on other author network
properties such as citation information. These co-authorship networks
are a special case of \emph{scientific collaboration
  networks}~\cite{moed2004handbook} and are a constant subject of
research, in particular with respect to scientometric analysis.  State
of the art studies investigate these networks in a global
scope~\cite{isfandyari2021global}, focusing on whole research
areas~\cite{coauthstatistics} or incorporate
temporal aspects~\cite{coauthstatistics}.

\paragraph{Topic Models}
At the current state of research, there is a variety of useful and
widely applicable topic models for use on text corpora.  The majority
of approaches to topic modeling take a document-word matrix as input,
in which documents are represented in a so called vector space
model. The first prominent instance is \emph{Latent Semantic
  Analysis}~\cite{deerwester90lsa}, short LSA, and is based on
the\footnote{This decomposition is unique up to the chosen basis.}
singular value decomposition of the input matrix. From the factor
matrices, one can infer relations between documents and topics as well
as topics and terms. A similar principle is followed by the
non-negative matrix factorization~\cite{Lee99MF} procedure (NMF),
which decomposes a matrix into two factors that are non-negative
matrices. This decomposition results in far better understandable
topic representation, since the topic values for a document cannot be
negative. The same is true for a topic's term values.

A probabilistic approach to topic modeling was introduced
by~\citet{lda} and is called \emph{Latent Dirichlet Allocation}
(LDA). This and thereof derived methods~\cite{dlda, blei_lafferty07,
  topicsovertime,bleiTopic} are broadly used in practice, however,
explaining them presents a certain obstacle. Moreover, when confronted with
short texts, such as research paper abstracts, the results of NMF are
superior to those of LDA~\cite{Hong2010}.

Since we want to base our investigations for \methodname{} on the
titles and abstracts from a large document corpus, we decided for
NMF. This decision was also influenced by the fact that LDA produces
considerably less stable results. This way, we can provide more
reproducible results in repeated runs compared to employing the LDA
algorithm~\cite{topicstability}. Finally, we refrain in this work from
using dynamic topic models and word embeddings as these also reduce
the explanatory power of our approach.

\paragraph{Topic evolution and Topic Flow}
The term \emph{topic flow} is still ambiguous in the scientific
literature. For example, in the area of online social network analysis
the authors~\citet{TwitterFlow} refer to \emph{TopicFlow} as a
visualization to capture the evolution of topics from discussions on
Twitter. In the realm of scientometrics, topic flow is often taken as
the share of a topic in the total amount of all scientific
publications in a year.  Basically, the research approach to date can
be divided into two categories:
\begin{inparaenum}
\item Intrinsic network-based modeling of topics and (potentially)
  their propagation in time (IN);
  \item Extending (temporal) networks by means of a topic model (EN). 
\end{inparaenum}
Yet, the concept of network in these categories ranges from
intradocument relations to interdocument clusters.

The work by \citet{churchill2018temporal} is an example for IN, which
employs a graph theoretic temporal topic model that identifies topics
as they emerge and tracks them through time.  The authors of
\citet{jiang2016text} propose a hierarchical topic model, an IN
approach, to capture the topic evolution over time. In a case study on
three computer science journals, the authors proposed a principal way
to visualize this topic evolution using Sankey diagrams. In contrast
to our work, however, the collaboration structure of the publication
network and the individual expertise of the authors were not taken
into account in the authors' approach.  A similar distinction applies
to ~\citet{li2019galex}, though in their paper the authors considered
topical co-authorship to be a relevant variable.

The second EN approach, linking
co-author networks with an underlying (or external) topic model, has
already been tried a few times.
Most related to our method is the work
by~\citet{AuthortopicsJeongLGC20}, whose objective is to capture
temporal patterns of research interests of authors over time. The
authors studied their approach on a (comparatively) rather small data
set of about 800 documents and they have not yet defined or measured
the flow between (research) topics.  Another example is the work of \citet{KakenDBLP}, 
who combined an LDA model to enhance an unsupervised link prediction
learning task involving a Japanese research database and the Digital
Bibliography \& Library Project.

Of particular relevance to the present work are the articles based on
\emph{the map equation}
by~\citet{infomap,rosvall2009map,rosvall2010mapping}, a statistical
approach to network analysis which does also incorporate topical
aspects. Although their work seems similar to ours, their approach is
based on a fundamentally different question: How can one capture
research fields (or their topics) using citation patterns? This method
is orthogonal to ours, which attempts to capture the topical research
flow (or knowledge flow) between individuals, and in an aggregated form,
between research fields themselves.

Finally, all studies cited in this section have in common that their
methods were not applied to publication corpora of comparable (large)
magnitude compared to the present work, cf.~\cref{table:mathcsdata}.

\section{Problem Description}\label{problem}
An elementary component of scientific work is the exchange of
knowledge on various research topics in the form of author
collaborations. This interchange within co-authorships generates a
flow of (topical) knowledge between the authors and therefore of their
respective research fields, which we refer to as \methodname\
(\methodshort). Understanding this flow of information on the research
topics is crucial to comprehend scientific advances over
time. Investigating \methodshort\ is a difficult problem, since (P1)
papers are comprised of many different research topics. An author can
be associated with the topics of his or her papers. Thus, as the paper
topics change so will the associated research topics of the author
(P2). Furthermore, (P3) author collaborations can take place within a
research field (intratopic) and between different research fields
(intertopic). Another challenge is the delay in the publication
process (P4), since a certain time passes from the creation of a
research work to its eventual publication. Finally, (P5) the direction
of topic flow depends on the relative expertise of the concerned
authors.

Analyzing \methodshort\ demands for a sophisticated network
structure that captures author collaborations and their research
topics over time. With our work, we propose the \tfn\ (\tfns) which
fulfills the requirements above and allows for further
analyses. The creation of the \tfns\ requires for automatic methods to
identify research topics in scientific corpora. This is a challenging
task by itself, since related topics may overlap and are not
distinctly separable.

\section{The \tfn}
\begin{figure}
\centering
\includegraphics[width=0.9\columnwidth]{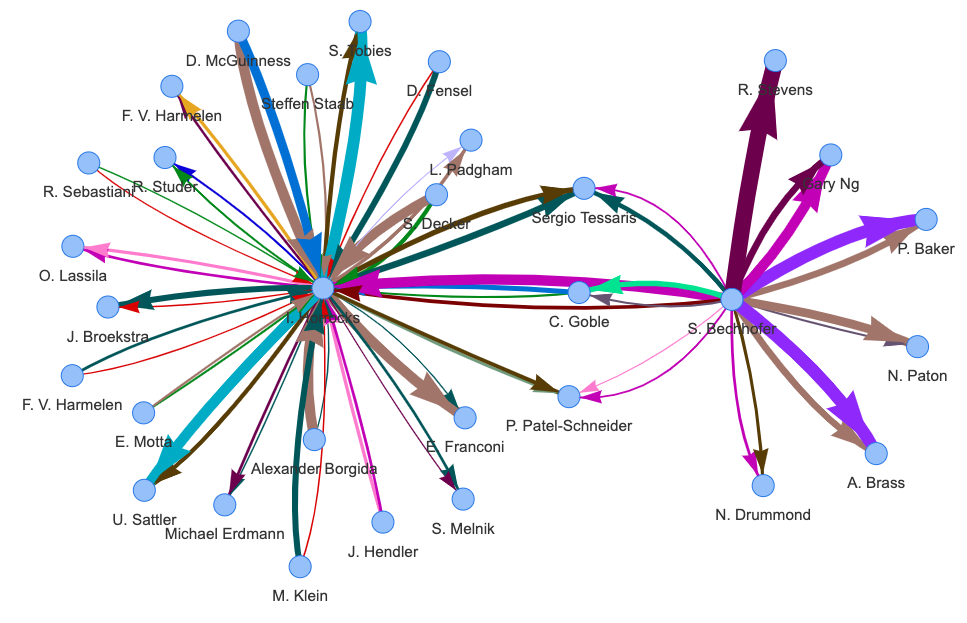}
\caption{\tfn{} example of the year 2000 in the neighborhood of the computer science researcher \emph{Ian Horrocks}. Edges between author nodes indicate topic flows. Edge colors represent topics and thickness indicates edge weight. For improved comprehensibility, the figure shows only a sample of the edges between the given nodes.}\label{fig:tfn}
\end{figure}
The construction of a \tfn{} is based on a \emph{publication corpus}
$C$, which is a relation of authors $A$, their papers $P$ in their
publication years~$Y$, in short, $C \subseteq P \times \mathcal{P}(A)
\times Y$. We will often select the works by an author $a\in A$
published in the year $y\in Y$ using the selection map $\sigma:A\times
Y \to \mathcal{P}(P)$,\linebreak $\sigma(a,y)\mapsto\{p\in P\mid
\exists (p,N,y)\in C\ \text{with}\ a\in N\}$. Due to the common delays
in the scientific publication process, we will in practice relax this
definition by additionally considering tuples $(a,N,y-1)$ and
$(a,N,y-2)$, see P4 in~\cref{problem}.

In order to measure (research) topic flow a model for topics on $C$ is
required. Given such a model $T$ using $|T|=n$ topics, we can derive a
map $\theta:P \to [0,1]^{\mid T\mid}, p\mapsto \theta(p)\coloneqq
(t_1,\dots, t_{n})$. The $n$ components of the topic vector reflect
the proportions to which extent each topic $t_{i}$ belongs to paper
$p$. This representation addresses P1 in~\cref{problem}. Whenever we
want to address a particular topic $t$ from a topic vector, we project
on it using $\pi_{t}$, e.g., $\pi_{t}\theta(p)$. By abuse of notation,
we employ the same function symbol for the map $\theta:A\times Y,
(a,y)\mapsto\theta(a,y)\coloneqq \sum_{p\in \sigma(a,y)}\theta(p)$, which
is the topic vector of an author $a$ in year $y$. Since
the arity of this function is different, we assume that there is no
risk of confusion.

\begin{definition}[\tfn]\label{def:tfn}
  A \emph{\tfn}(\tfns) is an edge-weighted multi-graph
  $G_{\theta}\coloneqq (A, \mathcal{E}_{\theta})$ which consists of an
  author set $A$ and a set of edge relations $\mathcal{E}_{\theta}\coloneqq\{E_{y,t}^{\theta}\}_{y\in
    Y,t\in T}$ with
  \[E_{y,t}^{\theta}\coloneqq \{(a_1,a_2)\in A\times A \mid
  \sigma(a_1,y)\cap \sigma(a_2,y) \neq \emptyset \wedge \pi_{t}\theta(a_1,y)\geq
  \pi_{t}\theta(a_2,y)\}\]
and the weight functions
\[\omega_{y,t}^{\theta}:E_{y,t}^{\theta}\to \mathbb{R}_{\geq0}, (a_{1},a_{2})\mapsto\omega_{y,t}^{\theta}(a_1,a_2)\coloneqq \pi_{t}\sum\nolimits_{p\in \sigma(a_1,y)\cap
  \sigma(a_2,y)}\theta(p).\]
 
\end{definition}

For the year $y$, authors $a_{1}$ and $a_{2}$ have a $t$-labeled edge
in $E_{y,t}^{\theta}$ if they published together on topic $t$ in this
year. For all practical purposes, we remind the reader of our
relaxation that will also consider the years $y-1$ and $y-2$.  The
weight for an edge in $E_{y,t}^{\theta}$ is the sum of topic vectors
of the publications written by $a_{1}$ together with $a_{2}$ projected on
topic $t$. We may stress that loop edges are included by this modeling. In
fact, we consider the weight of a self-loop of author $a$ on topic $t$ in
year $y$ to be the \emph{expertise of $a$ on $t$ in $y$}.

With this construction of the \tfns{} we want edges to reflect the
amount and the direction of topic flow within the co-authorship
network (see P5 in~\cref{problem}) as well as its change over time
(see P2 in~\cref{problem}). The direction is a direct result of
comparing topic vectors of the involved authors on the topic in
question. In our model we assume that topical knowledge flows from the
author with higher expertise to the one with lower expertise weight.
Different edges (i.e., different topics) between the same author pair
may have opposite directions in the same year. Moreover, our modeling
accounts for inter- as well as intratopic flow (see P3
in~\cref{problem}).

An example for a computed \tfn{} is given in \cref{fig:tfn}. In this figure, the \tfns{} of the year 2000 in the neighborhood of the computer science author \emph{Ian Horrocks} is depicted. Only a sample of the edges is shown for improved comprehensibility. Edges indicate topic flows between author nodes. Multiple edges on different topics, indicated by the colors, can occur between authors. Flows can have different weights, which is visualized by edge thickness. The data is taken from the case study in~\cref{sec:flowcasestudy}.


\subsection{\tfn\ Computation}
\paragraph{Corpus Preprocessing}
Computing topics for the construction of the \tfn{} requires that the papers $P$ in the publication corpus $C$ are converted to a vector space representation. For this, we concatenate the title and abstract of a paper, follow standard preprocessing techniques, such as tokenization and stop word removal and, finally, compute tf-idf representations~\cite{tfidf}. We may note, that preprocessing steps may be specific to the constitution of the input corpus rather than being an integral part of our overall method. For the more intricate details that may be involved in this process, we thus refer the reader to our case study in \cref{sec:flowcasestudy}.

\paragraph{Topic Model}
For the computation of (research) topics, we employ non-negative matrix factorization (NMF). From the tf-idf representation of the input corpus, NMF computes a given number of $n \in \mathbb{N}$ topics. Each computed topic is represented as a vector of weights, indicating relevances of terms to the topic. A topic can thus be interpreted by means of its top-weighted terms.
NMF additionally computes a topic representation of the input papers by means of weights, which indicate the relevance of each topic to each paper. Through this, we are able to construct the map $\theta$ for our \tfns. 

Since both, topic, and term weights, are non-negative, NMF gives topic
and paper representations that are well interpretable. In previous
research, the topics computed by NMF on scientific papers were found
to coincide with research topics~\cite{schaefermeier2020topic}. When
computing topics on a large publication corpus, a large number of
topics might be required to capture the different scientific
subfields. A challenge here is that the many topics of papers and the
resulting edges in a \tfns{} are difficult to comprehend, due to their
large number.

\subsection{Selection of Relevant Topics}
\label{sec:maintopics-flow}
In order to derive human-comprehensible knowledge from \tfns{} we
restrict the number of topics by removing certain edges from the graph
as well as defining main topics for nodes. For any author $a\in A$, we
consider the \emph{main topic in year $y$} to be the topic $t$
for which $a$ has the highest expertise in this year. We refer to this
topic using the partial function

\begin{equation}
  \label{eq:expertise}
  \tau:A\times Y\to T, \tau(a,y)\mapsto
  t=\argmax_{\substack{t\in T}}\omega_{y,t}^{\theta}(a,a).  
\end{equation}

Apart from being a partial function, \cref{eq:expertise} lacks
well-definedness with respect to the existence of a unique
maximum. This can be addressed by randomizing the selection. However,
in practice, i.e., when using sufficiently large topic models, these
exceptional cases are most-probably not encountered.

Whenever of interest, we will also refer to the second highest
weighted topic of an author $a$. This can be derived by
restricting~\cref{eq:expertise} to subsets of $T$.
Moreover, this approach allows for associating a main topic to any
subset $S\subseteq A$ of authors. The main topic $S$ is the most
frequent main topic in $S$, if existent. We acknowledge that this
approach may not work for very small or random sets of
authors. However, in real-world data sets we observe that this
definition is distinctive. Furthermore, we restrict the edge relations of
$G_{\theta}$  to the top-$l$ weighted edges per pair $(a_{1},a_{2})$
with $a_{1}\neq a_{2}$. For a suitable choice for $l$ we refer the
reader to the case study in~\cref{sec:flowcasestudy}.

\subsection{PageRank}\label{sec:pagerank}
PageRank~\cite{pagerank} is an algorithm to compute node relevances in
directed graphs. The assumption of PageRank is that nodes are relevant
when they have many incoming edges from other relevant nodes. The
PageRank algorithm repeatedly assigns relevance weights to nodes,
given some initialization, which is often the uniform weighting. In
every step, these weights are distributed evenly across the outgoing
edges of the nodes, i.e., passed to their neighbors. This process is
repeated until (some notion of) stability of the weights. Since we
figure that in our modeling topic weights should flow from lower to
higher topical expertise, we flip the edge directions in a \tfns{}
before computing PageRanks. Overall, this method allows us to
calculate importance weights for researchers within \tfns{} in a given
year. Furthermore, by restricting the \tfns{} $G_{\theta}$ to edges on
a certain topic $t\in T$, we are able to calculate relevances of
authors with regard to this given research topic $t$. Since PageRank
is a well-researched algorithm that has been proven in practice, it
was our first choice for analyzing \tfns{}. Yet, naturally, the whole
tool-set of centrality measures for directed graphs may be applied to
\tfns{} in future work.

\subsection{Community Detection}\label{sec:communitydetection}
One particular analysis that we want to carry out on topic flow
networks is \emph{community detection}. The general task of community
detection is to find densely connected subgraphs. Given a \tfn{}
$G_{\theta}$ this idea enables us to find research communities
consisting of collaborating authors of $A$. For community detection we
use the \emph{Walktrap} algorithm~\cite{walktrap}, which computes a
partition of the input graph node set $A$. The method is based on the
principle that a set of generated random walks within the graph tends
to get ``trapped'' inside the same, densely connected parts of this
graph. Walktrap has a comparatively low run-time complexity of $O(n^2
\log n)$, with $n=|A|$, in sparse as well as dense graphs. We chose
Walktrap for its resilience with respect to a wide range of network
characteristics, in particular the distinctiveness and fuzzyness of
communities~\cite{waltktrapIsResilient}.

We compute the partitions of the author nodes $A$ in $G_{\theta}$ for
each year $y \in Y$. To obtain some interpretation of the found
communities, we analyze the main topics $\tau(a,y)$ of their authors
$a$ as described in~\cref{sec:maintopics-flow}. Hence, we call the
most frequent topic within a community (i.e., a set of authors)
\emph{the main topic of the community}. Since \tfns{} can be very
large, e.g., as investigated in~\cref{sec:flowcasestudy}, the obtained
number of communities can be large as well. Therefore, we will
introduce aggregations and summary statistics in the
practical study. 

\subsection{$k$-Cores}\label{sec:kcores}
The $k$-core of graph $G=(V,E)$ is the maximal induced subgraph
$G[V']$ s.t. all vertices in $G[V']$ have at least node degree
$k$. Based on this the \emph{core number} of a vertex $u\in V$ is the
largest number $k$ s.t. $u$ belongs to the $k$-core. The largest core
number of a node in a graph is called the \emph{coreness} of $G$. We
compute the coreness for all subgraphs of $G_{\theta}$ that are
induced by the topics $t\in T$ and years $y\in Y$. In these settings,
we lift the definition of $k$-cores to multi-graphs where the degree
of a vertex is the sum of the individual degrees with respect to all
edge relations. Edges with weight zero are not considered.  We want to
remind the reader that a collaboration of authors $N\subseteq A$ leads
to multiple edges for every pair of authors in $N$
(\cref{def:tfn}). This modeling results in comparatively larger
core-numbers.

With $k$-cores we are able to assess the structural connectedness of
vertices with respect to the whole graph. This property is in contrast
to simple degree sequences, which only account for the neighborhood of
vertices. With the coreness of a topic subgraph in $G_{\theta}$ we
appraise the extent of research networking taking place through
collaboration on a given topic in a given year.

\subsection{Intra- and Intertopic Flows}\label{section:intertopicflows}
The main objective of the present work is to develop and analyze the
concept of intra- and intertopic flows between research topics. Both
can be modeled and captured by means of the \tfn{} in the following
way. We simplify our view on the graph $G_{\theta}$ by mapping to each
author his or her main topic (\cref{sec:maintopics-flow}) in year
$y$. This enables a simple clustering of all author nodes in this year
where any two elements of a cluster share the same main topic. More
formally, given the set of topics $T$ we find an equivalence relation
$\sim_{y}$ on $A$ by $a_{1}\sim_{y} a_{2}\colon\logeq
\tau(a_{1},y)=\tau(a_{2},y)$ and the corresponding clustering
(partition) is denoted by $A_{/\sim_{y}}$. This partition, in turn,
allows for computing a topic flow between any two topics.

\begin{definition}[Topic Flow]
  For \tfns{} $G_{\theta}$ with topics $T$ let $A_{1},A_{2}\in
  A/_{\sim_{y}}$, where $t_{1}\in T$ is the main topic for all $a\in
  A_{1}$ and $t_{2}\in T$ is the main topic for all $b\in A_{2}$. The
  \emph{topic flow from $t_1$ to $t_{2}$} is
  \[\phi_{y}(t_{1},t_{2})\coloneqq\sum_{\substack{a\in A_{1}\\ b\in A_{2}}} \omega_{y,t_{1}}^{\theta}(a,b).\]
\end{definition}

This definition defines the intra- and intertopic flow between any two
(different) research topics within a \tfns{}. It is based on the
assumption that such a topic flow arises between any two authors from
(different) research fields (i.e., with different main topics) when
they collaborate.  More specifically, two authors $a$ and $b$ with
main topics $t_1$ and $t_2$ contribute to the intertopic flow from
$t_1$ to $t_2$ with the weight of the edge $(a,b)$ on topic $t_1$. The
sum of all such contributions is the topic flow from $t_{1}$ to
$t_{2}$. We refer to any flow $\phi_{y}(t,t)$ as an \emph{intratopic
  flow} and accordingly any flow $\phi_{y}(t_{1},t_{2})$ with
$t_{1}\neq t_{2}$ as an \emph{intertopic flow}. These flows allow us,
first, to capture cross-topic collaborations in general, and second,
to quantify the extent of such collaborations. With the latter we assume
to measure in particular the flow of (topical) expertise from $t_{1}$
to $t_{2}$, which may influence the target topic $t_{2}$.

\section{\methodname\ in Math and Computer Science}
\label{sec:flowcasestudy}
In order to test and evaluate \tfn{} we conducted a case study on two
comprehensive publication corpora $C_{\text{MATH}}$ and $C_{\text{CS}}$ from the research fields
\emph{Mathematics} and \emph{Computer Science}. We compiled the data
basis by extracting publications from the Semantic Scholar Open
Research Corpus (S2ORC)~\cite{semanticscholar}. This extraction was
constrained to publications that were designated either Mathematics or
Computer Science in the attribute \emph{fields of study} and were
published between 1960 and 2021.  
For the creation of the math corpus $C_{\text{MATH}}$, we solely used
publications that were marked as Mathematics and not as Computer
Science. This decision arised from the observation that papers marked
as both tend to focus on computer science. Based on both data sets, we
created topic flow networks as expounded in the previous section. In a
given year $y\in Y$, we restricted the number of edges per
collaboration of authors $a,b\in A$ to the top-8 topics in order to
remove topics with low contributions. This was done for two reasons:
First, the thus created graph can be analyzed efficiently through
algorithmic, in particular network centered, approaches.  Second,
bounding the number of topics results in a more human-comprehensible
analysis process.
%
\cref{table:mathcsdata} gives an overview on the created corpora and the resulting topic flow networks.

\begin{table}[h!]
\centering
\caption{Statistics for the Math ($C_{\text{MATH}}$) and the Computer
  Science ($C_{\text{CS}}$) corpora and the resulting topic flow
  network graphs.}
\begin{tabular}{r  | r | r} 
 \toprule
& \textbf{Math} & \textbf{Computer Science} \\
 \hline
  \emph{Publications} & 5,314,915 & 14,677,697 \\
  \emph{Publications with abstract} & 3,976,750 & 10,992,167  \\
  \emph{Authors (nodes)} & 185,835 & 714,212 \\
  \emph{Collaborations (edges)} & 126,693,634 & 736,891,384  \\ 
   \emph{Year range} & 1960 - 2021 & 1960 - 2021 \\
   \bottomrule
\end{tabular}
\label{table:mathcsdata}
\end{table}

\subsection{Topic Flow Network Computation}
\paragraph{Corpus Preprocessing}
For preprocessing, we concatenate titles and abstracts and tokenize
documents. Since we found many papers written in Indian, Chinese,
Japanese and Russian language, we remove non-English documents based
on a simple heuristic: If at least ten percent of the tokens in a
document are contained in an English stop word list,\footnote{Used
  stop word list: \url{https://github.com/RaRe-
    Technologies/gensim/blob/develop/
    gensim/parsing/preprocessing.py}} we classify a document as being
in English language. We determined the 10\% threshold through a manual
examination of publications with stop word proportions below different
thresholds.  We compared this heuristic to a more computationally
intensive approach, the Python \emph{langdetect}
package\footnote{\url{https://pypi.org/project/langdetect/}}, that is
widely used in practice. Assuming the results obtained by
\emph{langdetect} as ground truth values, our method resulted in an
F1-score of 0.993 on the computer science data set. We found this
outcome to be sufficiently close to \emph{langdetect} given that it
led to substantially reduced computation
time. 
About 5\% of the papers were removed through this process. As a next
step, we remove stop words based on the same list as above. We do
not use any stemming, since this may reduce terms with different
meanings to the same word stem, which would be especially problematic
for the recognition of topics in a scientific context. Finally, we
compute tf-idf representations for all publications~\cite{tfidf}.


For the topic model training, we solely  use papers having  an
abstract. We base this decision on the assumption that titles alone
can have negative effects on topic model training due to a different
distribution of the tf-idf values. This step removed about 25\% of the
documents. Yet, for all consecutive analyses we employ all documents,
i.e., also documents without an abstract. 

\paragraph{Topic Model}
A crucial parameter in the topic model training is the \emph{number of topics} to be found.
We experimented with different topic numbers on both, the computer science corpus $C_{\text{CS}}$ and the math corpus $C_{\text{MATH}}$. Based on a manual assessment of the obtained topical granularity, we finally decided for 64 topics in both data sets. We additionally based this decision on the number of topics in the Mathematics Subject Classification, which is in a similar range.\footnote{\url{https://mathscinet.ams.org/msc}} Moreover, we initially experimented with a coherence measure but found the resulting optimal topic number too low to reflect the variety of the research fields that is contained in a large, comprehensive publication corpus. We ensured convergence of the topic model training by visual inspection of the training error. For all other hyperparameters, we used the defaults from the \emph{gensim} library.\footnote{\url{https://radimrehurek.com/gensim/models/nmf.html}} The computed topics for both data sets are given in the appendix in \cref{table:mathtopics,table:cstopics}. These are represented as a list of their respective top five weighted terms in the NMF model. For example, we were able to derive important research fields from $C_{\text{MATH}}$, such as \emph{group theory} (Topic 16) and \emph{coding theory} (Topic 47). Similarly, in $C_{\text{CS}}$ we found topics such as \emph{neural networks} (Topic 42) and \emph{search engines} (Topic 10).

\subsection{PageRank}
We employ the PageRank algorithm, as described in~\cref{sec:pagerank}, to identify researchers that
stand out for their collaboration relationships in the \tfn. We
conduct this analysis in three settings. First, we compute PageRank in a \tfns{} representation of $C_{\text{MATH}}$. Second we proceed in the same way for the $C_{\text{CS}}$ corpus. Third, we restrict the \tfns{} from $C_{\text{CS}}$ to $t=$\,\emph{robotics} (Topic 26), see
\cref{table:pagerankcs,table:pagerankmath}.

When comparing the ranked researchers to common scores, such as citation count and h-index\footnote{These were extracted from Semantic Scholar in mid 2022.}, we find that the highly ranked authors score high on average. In the robotics field, our method identified, e.g., S. Thrun, a well-known researcher in the field, as a top ten ranked author. 
The other authors in this ranking are also established researchers in the fields of robotics, as an empirical review of their publications reveals. Within $C_{\text{MATH}}$, our method ranked Paul Erdős second. Since he is a prime example of a intertopical researcher in mathematics, we count this as a success of our approach. 
We want to stress that our PageRank approach differs
from statistics such as h-Index and citation counts in that it accounts for the
entire (topical) network structure.

Altogether, we find that topic flow networks are suitable for the
automatic discovery of important authors for a given research topic. Our approach is capable of
providing a momentary influence indicator for researchers in a given time window or topic. Hence, in contrast to citation-based methods, our approach can reveal recent intertopical flow and its (most relevant) generating authors. This in particular true in research fields where the citation frequency is low, e.g., mathematics.

\begin{table}
\centering
\caption{Top ten ranked authors computed by PageRank in $C_{\text{CS}}$ for the year 2000. *Note: H-Index values and citation counts ($n_\text{cite}$) were extracted from the Semantic Scholar website in mid 2022.}
\begin{tabular}{r| l r r  |  l  r r}
\toprule
&  \multicolumn{3}{l}{\textbf{Computer Science (2000)}} & \multicolumn{3}{l}{\textbf{Robotics (2000)}} \\
\hline
\emph{Rank} & \emph{Author} & $n_{\text{cite}}$* & \emph{H-Ind,*} & \emph{Author} &  $n_{\text{cite}}$* & \emph{H-Ind.*} \\
\toprule
\emph{1} & T. Fukuda & 26,104 & 70 & T. Fukuda & 26,104 & 70 \\
\emph{2} & T. S. Huang & 103,000 & 150 & D. Thalmann & 23,985 & 80 \\
\emph{3} & A. Sangiovanni & 49,353 & 101 & H. Kitano & 6,431 & 38 \\
\emph{4} & J. Kittler & 51,970 & 89 & M. Asada & 11,856 & 46 \\
\emph{5} & C. Suen & 26,218 & 66 & M. Veloso & 32,132 & 74 \\
\emph{6} & T. Kanade & 99,184 & 140 & R. Simmons & 17,719 & 67 \\
\emph{7} & F. Catthoor & 17,239 & 56 & H. Asama & 7,287 & 36 \\
\emph{8} & J. Dongarra & 65,552 & 114 & G. Hirzinger & 28,737 & 83 \\
\emph{9} & C.-C. Jay Kuo & 29,491 & 77 & S. Thrun & 97,400 & 145 \\
\emph{10} & R. Kikinis & 61,726 & 126 & K. Tanie & 10,270 & 48 \\
\bottomrule
\end{tabular}\label{table:pagerankcs}
\end{table}

\begin{table}
\centering
\caption{Top ten ranked authors computed by PageRank in $C_{\text{MATH}}$. *Note: H-Index values and citation counts  ($n_\text{cite}$) were extracted from the Semantic Scholar website in mid 2022.}
\begin{tabular}{r| l r r  |  l  r r}
\toprule
&  \multicolumn{3}{l}{\textbf{Math (1965)}} & \multicolumn{3}{l}{\textbf{Math (2020)}} \\
\hline
\emph{Rank} & \emph{Author} &  $n_\text{cite}$ & \emph{H-Ind.*} & \emph{Author} & $n_\text{cite}$ & \emph{H-Ind.*} \\
\toprule
\emph{1} & R. Bellman & 62,117 & 84 & D. Baleanu & 48,095 & 93 \\
\emph{2} & P. Erdős & 38,540 & 93 & H. Srivastava & 46,826 & 79 \\
\emph{3} & D. Speiser & 737 & 12 & K. Nisar & 5,897 & 35 \\
\emph{4} & E. Robinson & 4,835 & 29 & P. Kumam & 11,501 & 44 \\
\emph{5} & R. Kalaba & 11,465 & 47 & T. Abdeljawad & 10,536 & 52 \\
\emph{6} & S. Karlin & 42,178 & 93 & Y. Chu & 12,308 & 53 \\
\emph{7} & H. Davenport & 8,853 & 42 & D. O’Regan & 22,240 & 64 \\
\emph{8} & K. Parthasarathy & 6,524 & 29 & F. Smarandache & 23,969 & 67 \\
\emph{9} & A. Green & 15,901 & 53 & R. Agarwal & 36,790 & 83 \\
\emph{10} & O. Kempthorne & 8,642 & 46 & A. Alsaedi & 58,137 & 100 \\
\bottomrule
\end{tabular}\label{table:pagerankmath}
\end{table}

\subsection{Social Network Analysis}

\begin{figure}
\includegraphics[width=0.5\columnwidth,trim={0.4cm 0.4cm 0.4cm 0.4cm},clip]{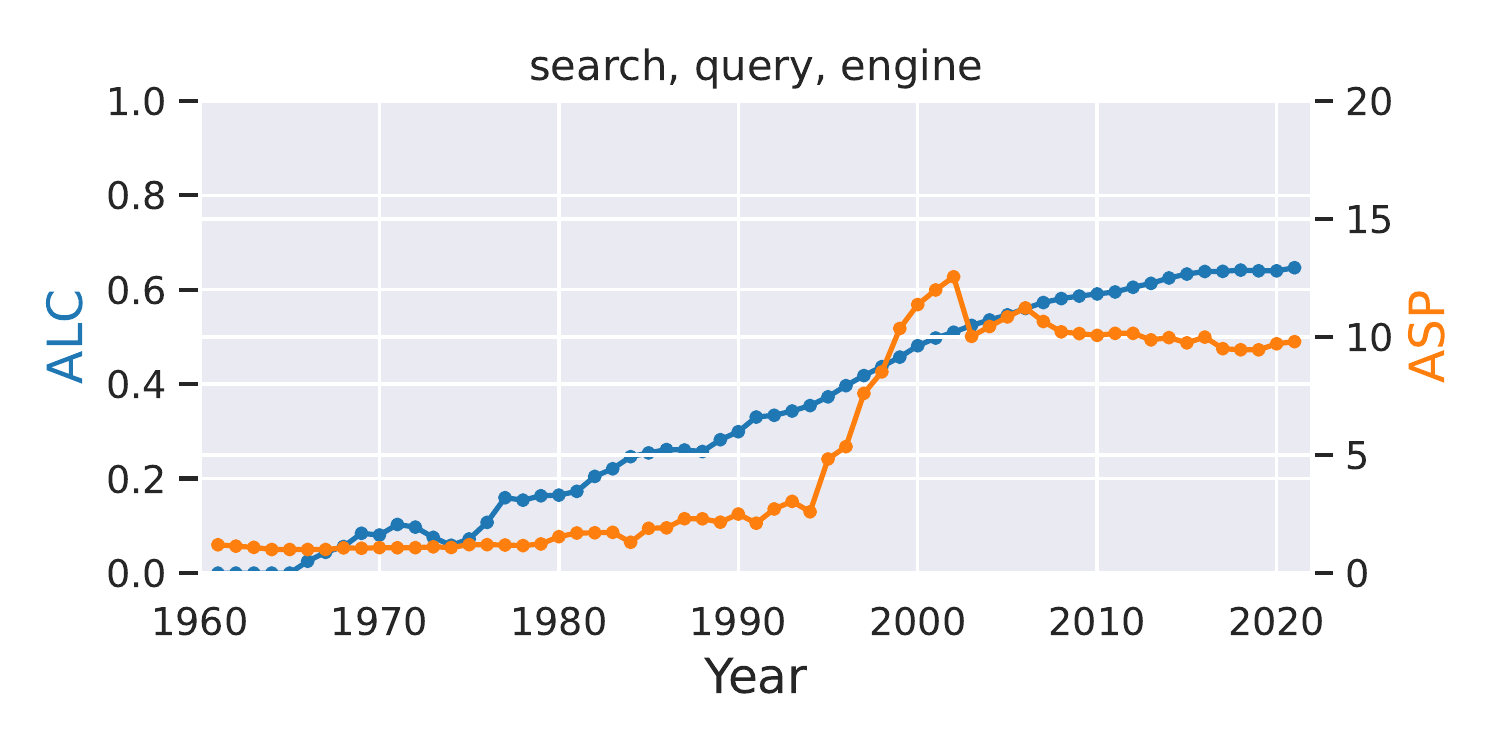}
\includegraphics[width=0.5\columnwidth,trim={0.4cm 0.4cm 0.4cm 0.4cm},clip]{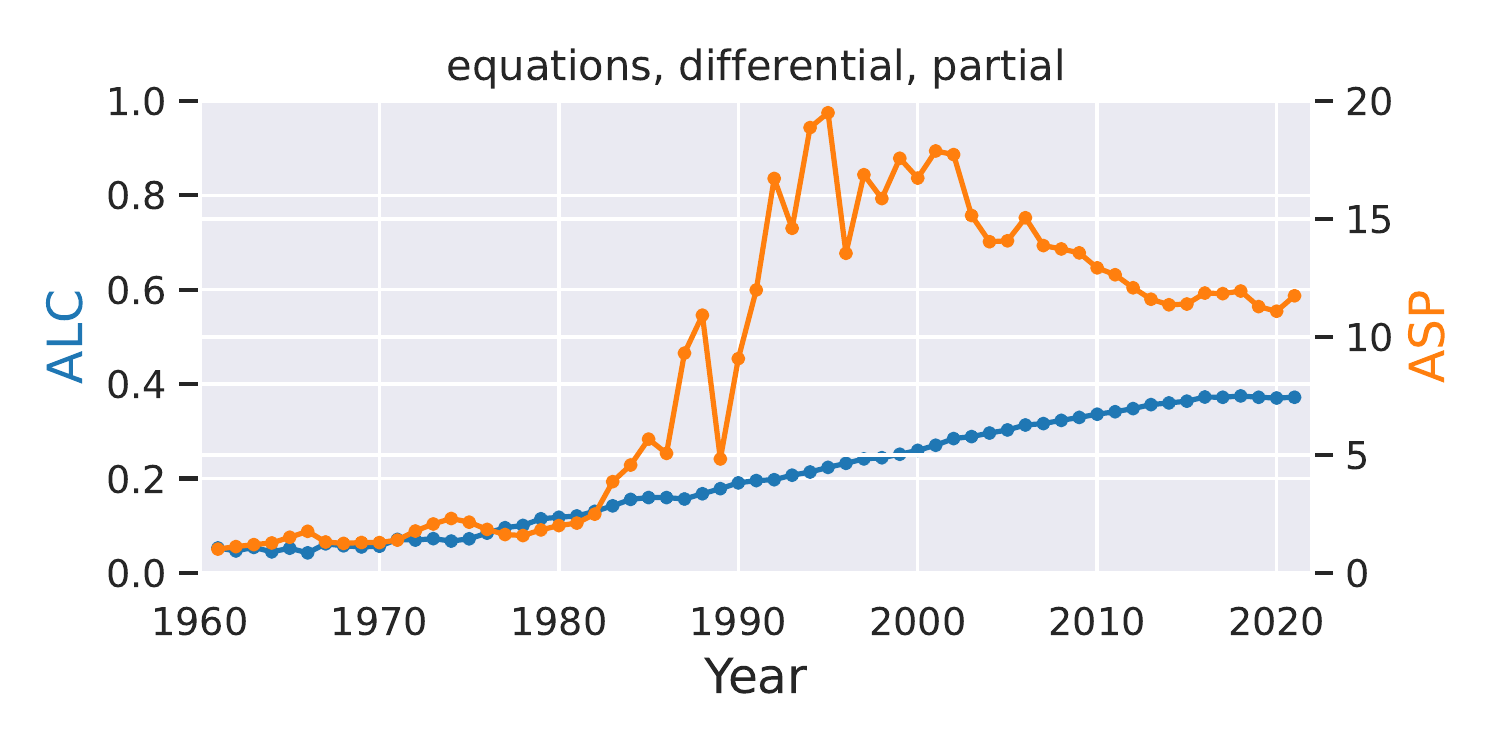}
\includegraphics[width=0.5\columnwidth,trim={0.4cm 0.4cm 0.4cm 0.4cm},clip]{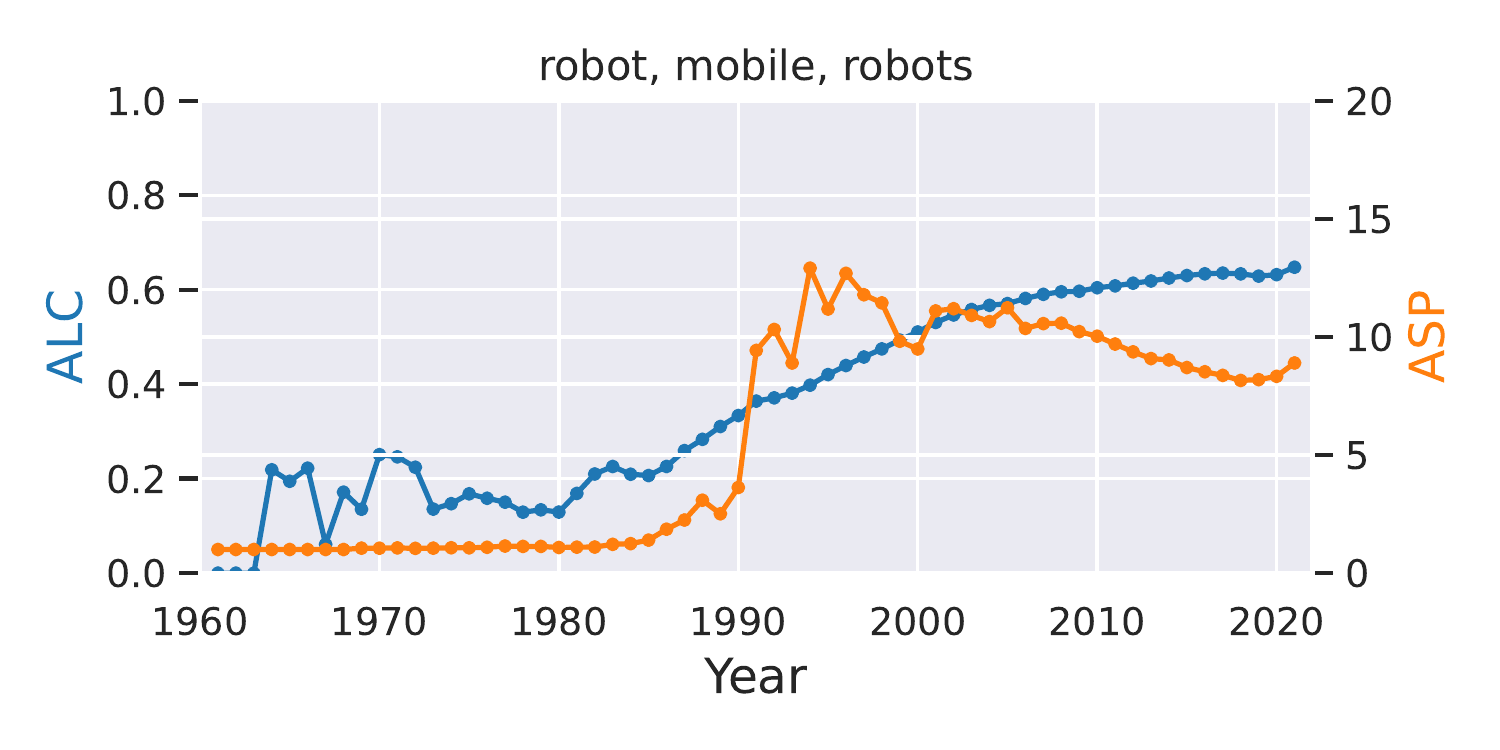}
\includegraphics[width=0.5\columnwidth,trim={0.4cm 0.4cm 0.4cm 0.4cm},clip]{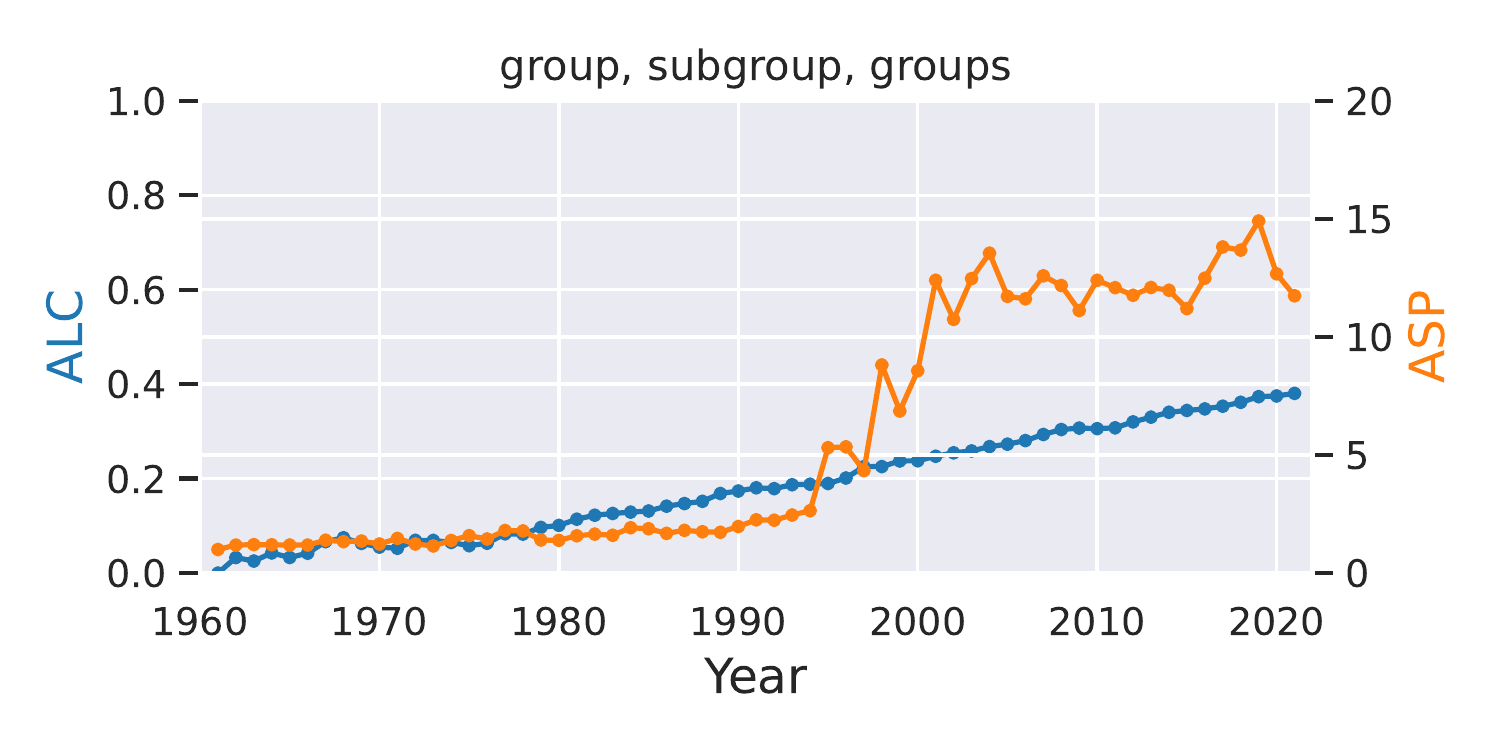}
\includegraphics[width=0.5\columnwidth,trim={0.4cm 0.4cm 0.4cm 0.4cm},clip]{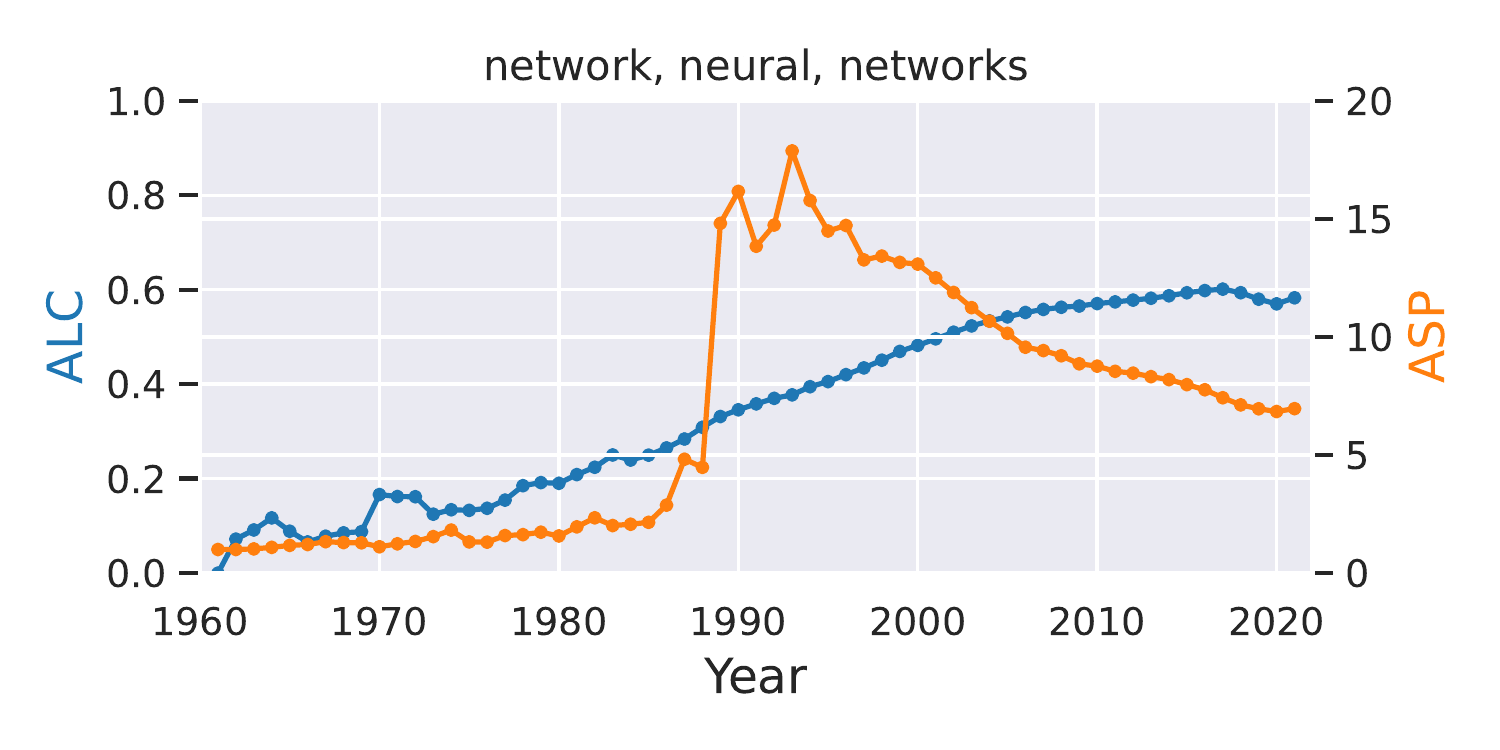}
\includegraphics[width=0.5\columnwidth,trim={0.4cm 0.4cm 0.4cm 0.4cm},clip]{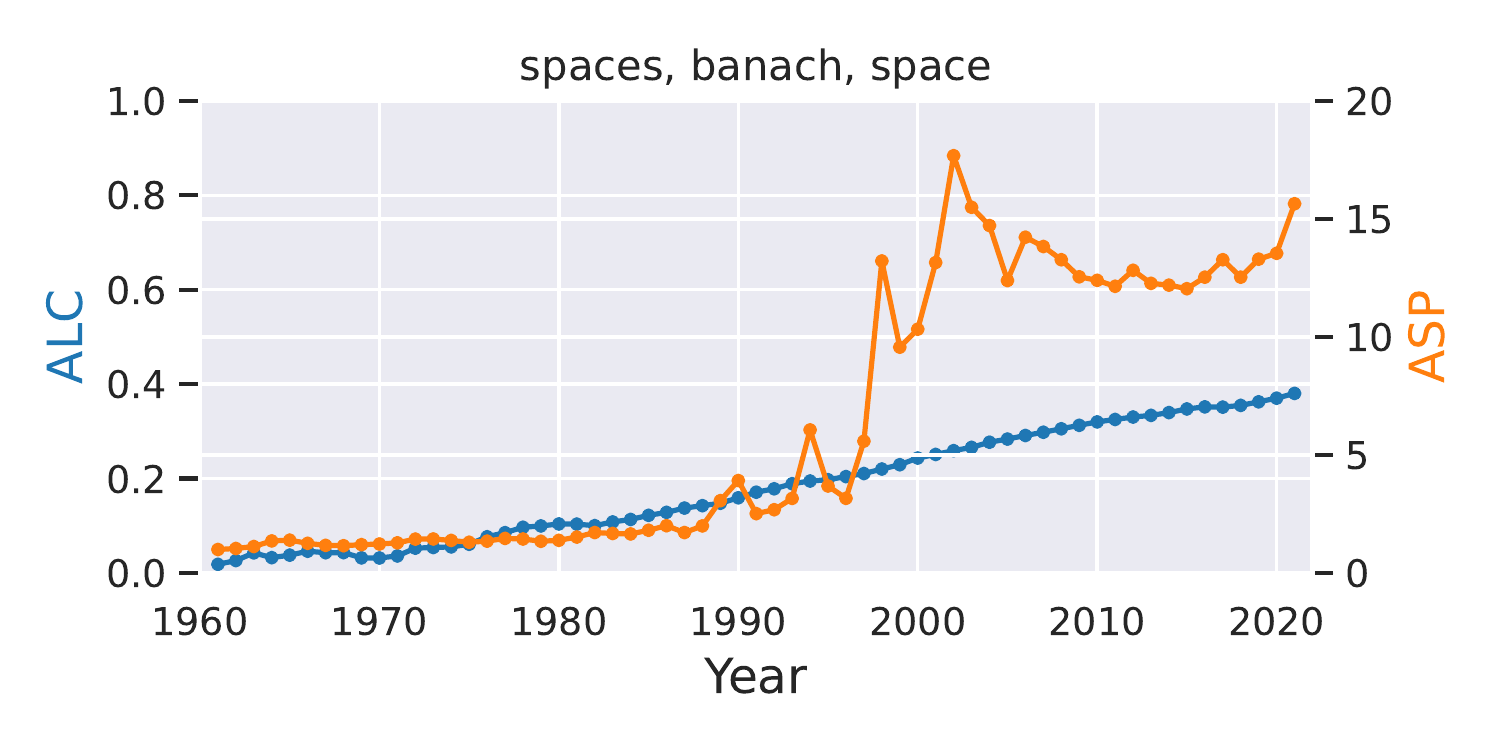}
\caption{Average shortest path (ASP) and average local clustering coefficient (ALC) for different math and computer science topics. Topics in the left column are from computer science and topics in the right were computed on the math corpus.}\label{fig:aspalc}
\end{figure}
Apart from identifying outstanding researchers in the \tfn,
we are interested in grasping the overall network structure.
Naturally, \tfns s are social networks and can therefore be treated as such, using the whole toolset of network analysis.
A particular question in this direction is to
which extent a \tfns{} satisfies the characteristics of a
\emph{small-world} network. For this analysis, we employ the most important network properties, \emph{average local clustering coefficient} (ALC) and 
\emph{average shortest path} (ASP). Both  metrics have been reported in the literature as relevant to the study of collaboration graphs~\cite{NewmanShortestPathEssentialCollab}. We computed these properties for three computer science and mathematics topics respectively. We did this for all years available in the corpora and depicted the results in \cref{fig:aspalc}. Notably, there is a growth of the ALC over years in all networks, which for most topics looks almost linear. This indicates an increasing local connectedness, i.e., collaboration, of researches. Furthermore, we find that most recent values for ALC appear to differ structurally between the research fields mathematics and computer science.

The ASP on the other hand has a sudden increase between 1990 and 2000 for all topics. For some topics, e.g., neural networks, we observe that a sudden peak is followed by a decrease. We surmise that the sudden increase of the ASP occurs due to the overall growth of the network, while the decrease indicates increasing connectedness, i.e., triadic closure.
We presume that a driver for this growth could be the global political change in the 1990s~\cite{MoreInternationalAfter1990} and the therewith lifted restrictions on international collaboration. Moreover, in particular for the topic \emph{search, query, engine}, we suspect the more wide-spread use of the internet and the thereof resulting need for search technologies (within not curated collections of data) may have had a substantial effect~\cite{sanderson2012historyOfInformationRet}. Altogether we conclude that \tfns{} grasped as social networks enable a variety of possibilities for topic-centered scientometric analyses.

\subsection{Community Detection}\label{sec:communityexperiment}
\begin{figure}
\includegraphics[width=\columnwidth,trim={0.9cm 1cm 5.5cm 2cm},clip]{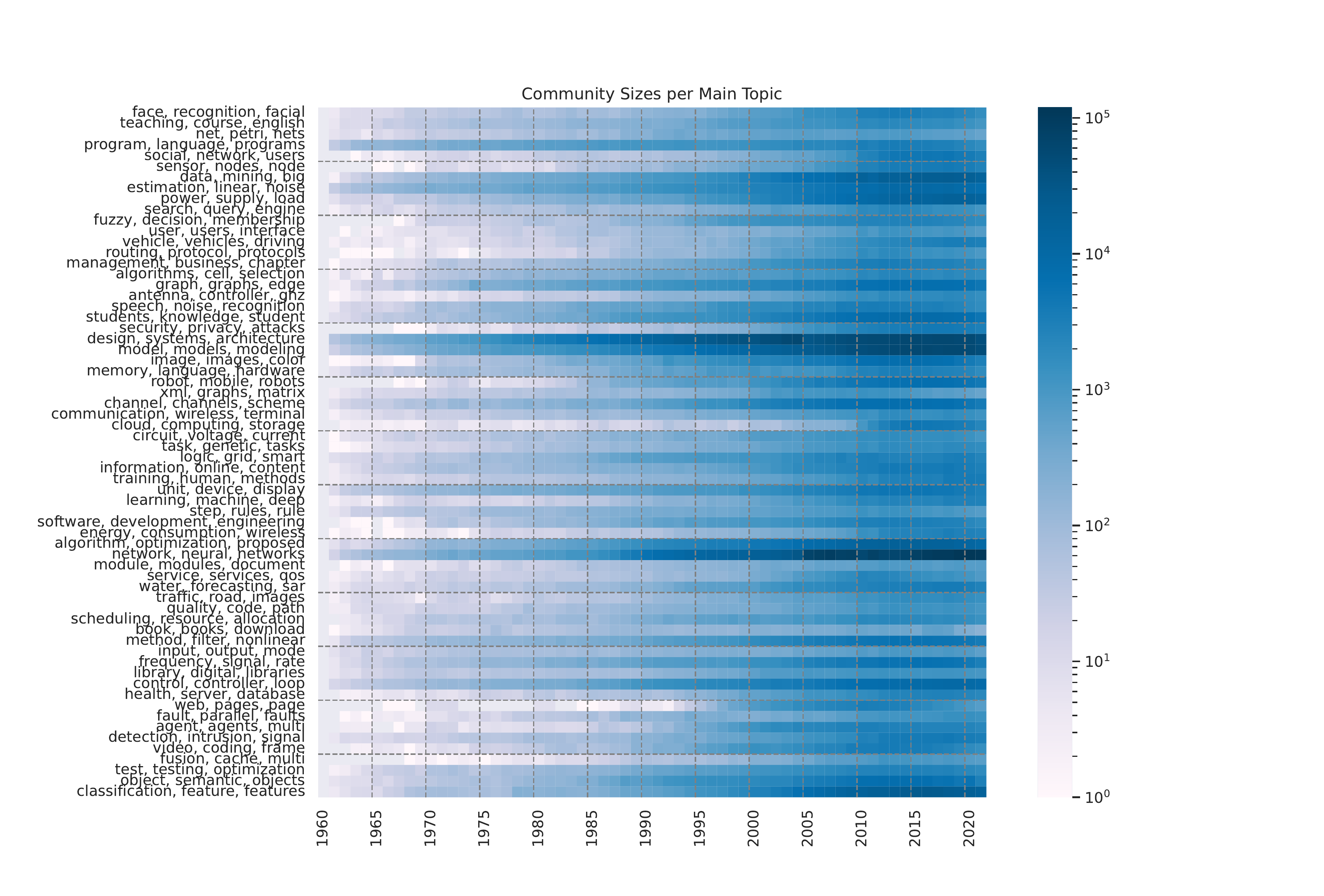}
\caption{Community sizes for computer science data set $C_{\text{CS}}$.}\label{fig:communitiescs}
\end{figure}
To reveal community structures in \tfn{} s and how they change
over time, we applied the
Walktrap algorithm with default
parameters (see \cref{sec:communitydetection}). For each year, we
applied this algorithm and obtained communities $O_i$ in the form of subsets of the authors $A$. For every $O_i$ we computed its size and the main topic of the contained authors. In the following, we omit all communities of size 1, i.e., isolated researchers. This modeling allows for the application of various community analysis methods.
As an example, for any topic $t$ we summed up the sizes of all communities with this main topic and depicted the results for $C_{\text{CS}}$ in  \cref{fig:communitiescs} and for $C_{\text{MATH}}$ in the appendix in~\cref{fig:communitiesmath}.

Investigating the $C_{\text{CS}}$ results,
we find that the number of communities increases for almost all topics over time. This may be a
consequence of the overall growth of the number of scientific
authors. Nonetheless, we can identify several topics for which the community sizes decrease after a certain point in time. Furthermore, we are able to identify the rising interest in certain
research topics. For example, the community sizes for the topic
\emph{web, page, pages} begin to grow in the late 1990s, which coincides with the
broad use of the web (9th row from bottom). 
Around 2015 interest in this topic decreased again, possibly due to a more
differentiated terminology and increasing research on, e.g., social networks
(\emph{social, network, users}, 5th row from top) and \emph{cloud, computing, storage} (30th row from top) ,
which gained interest around 2010.

For some examples, we looked into the two most frequent topics
for some communities (\cref{sec:maintopics-flow}). For
example, in 2021 the largest community $C_i$ had the two most frequent topics
\emph{network, neural, networks, layer, deep} and \emph{model, models,
  simulation, prediction}. For the second largest
community, we found the topics \emph{classification, feature,
  features, classifier, accuracy} and \emph{network, neural, networks,
  layer, deep}. As another
example, the fifth largest community, we found \emph{network, neural, networks,
  layer, deep} in combination with \emph{image, images, color,
  segmentation, processing}. In all these cases, both topics are strongly fitting semantically. We take this as evidence that the Walktrap algorithm detected meaningful communites. Moreover, we conclude that using several main topics may lead to more distinguished descriptions of communities.
Altogether, topic flow networks appear to be suitable for the detection of research communities. More elaborate approaches for their analyses would be possible, e.g., based on properties such as author countries and institutions.

\subsection{$k$-Cores}\label{core-experiment}
\begin{figure}
\includegraphics[width=\columnwidth,trim={0.9cm 1cm 5.5cm 2cm},clip]{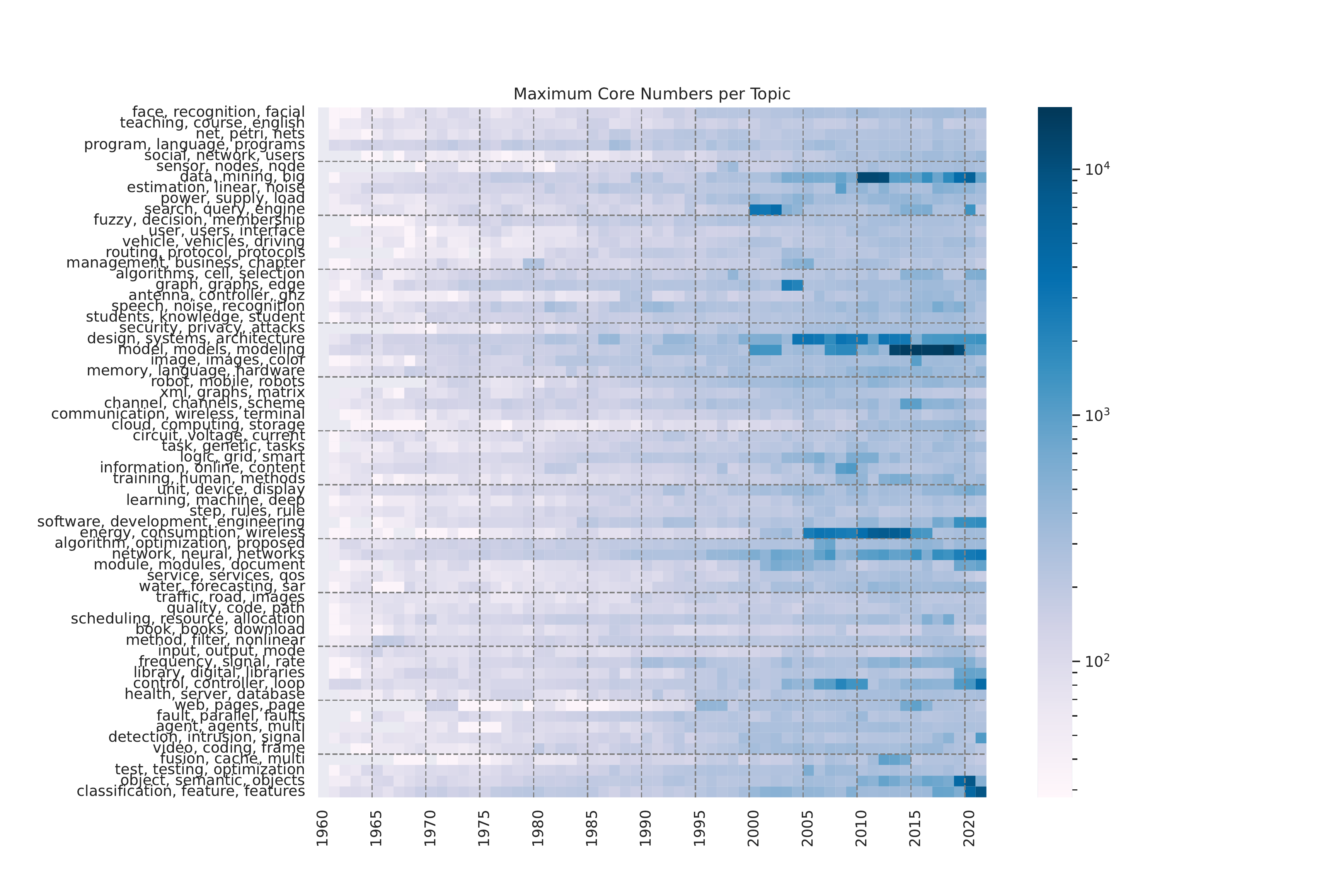}
\caption{Maximum core numbers for computer science data set $C_{\text{CS}}$.}\label{fig:corescs}
\end{figure}
We compute the coreness of
the $C_{\text{CS}}$ \tfns{} restricted to all topics $t \in T$ based on the approach explained in
\cref{sec:kcores}. With the coreness of a topic network in a year, we try to
assess the degree of networking that takes places through
collaboration. We depict in the heat map in \cref{fig:corescs} the coreness of all topics and all years considered. 
The color intensity reflects the computed values.
We added the respective results for  $C_{\text{MATH}}$ to the
appendix in \cref{fig:coresmath}. 

First, we observe that there is a substantial change in coreness
values beginning from around the year 2000. A general increase in coreness is expected as this
number is limited by the number of authors in the network and the therewith bounded number of edges. However, the sudden increase of coreness observed for several topics, such as \emph{data, mining, big} (7th row), \emph{energy, consumption, wireless} (40th row) and \emph{network, neural, networks} (42th row) shows that there exists some particularly strong collaboration by authors within these topics.
In detail, we find that the topic \emph{search,
  query, engine} (10th row) has a large
increase in coreness between 2000 and 2003, a time when the internet use spiked, and therefore the research question for finding information in it. 

\begin{figure}
\includegraphics[width=\columnwidth,trim={0cm 0cm 0cm 0cm},clip]{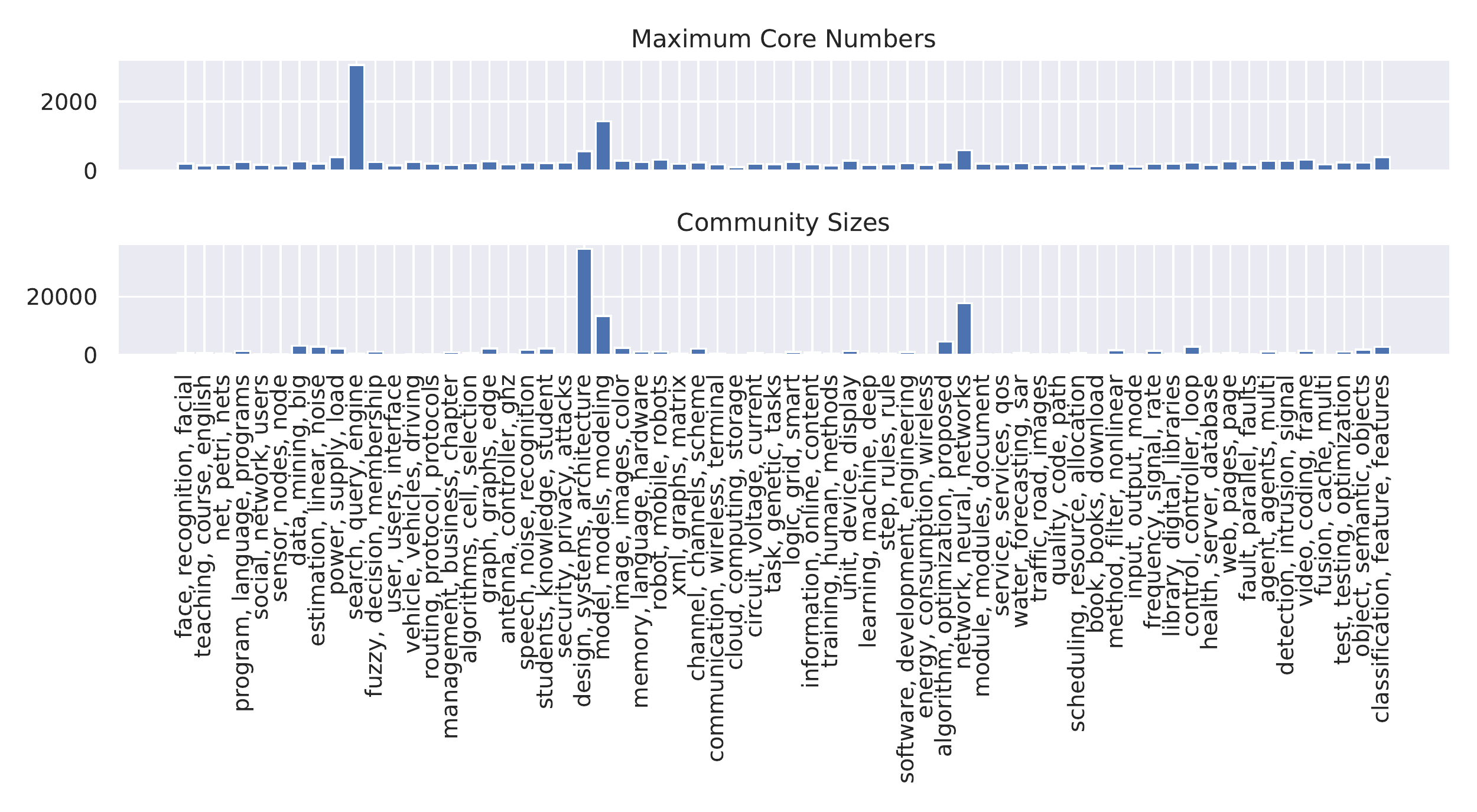}
\caption{Maximum core numbers for computer science data set in 2000.}\label{fig:cores2000}
\end{figure}

In \cref{fig:cores2000} we depicted the coreness values for all research topics in the year 2000. We contrasted these figures with the sum of the community sizes per topic from that year, as described in ~\cref{sec:communityexperiment}.
We notice that large community size does not necessarily imply large coreness and vice versa.
For example, there are few search engine communities and they are all comparatively small. However, the corresponding coreness is high, in fact the maximum observed value, which indicates that the search engine communities in that year are densely connected.
In summary, we conjecture that $k$-cores in \tfns s are capable of revealing new structural insights into publication corpora  relevant to scientometric analyses.

\subsection{Intra- und Intertopic Flows}
\begin{figure}
\centering
\begin{overpic}[width=0.9\columnwidth,trim={2cm 1.95cm 2cm 2.4cm},clip]{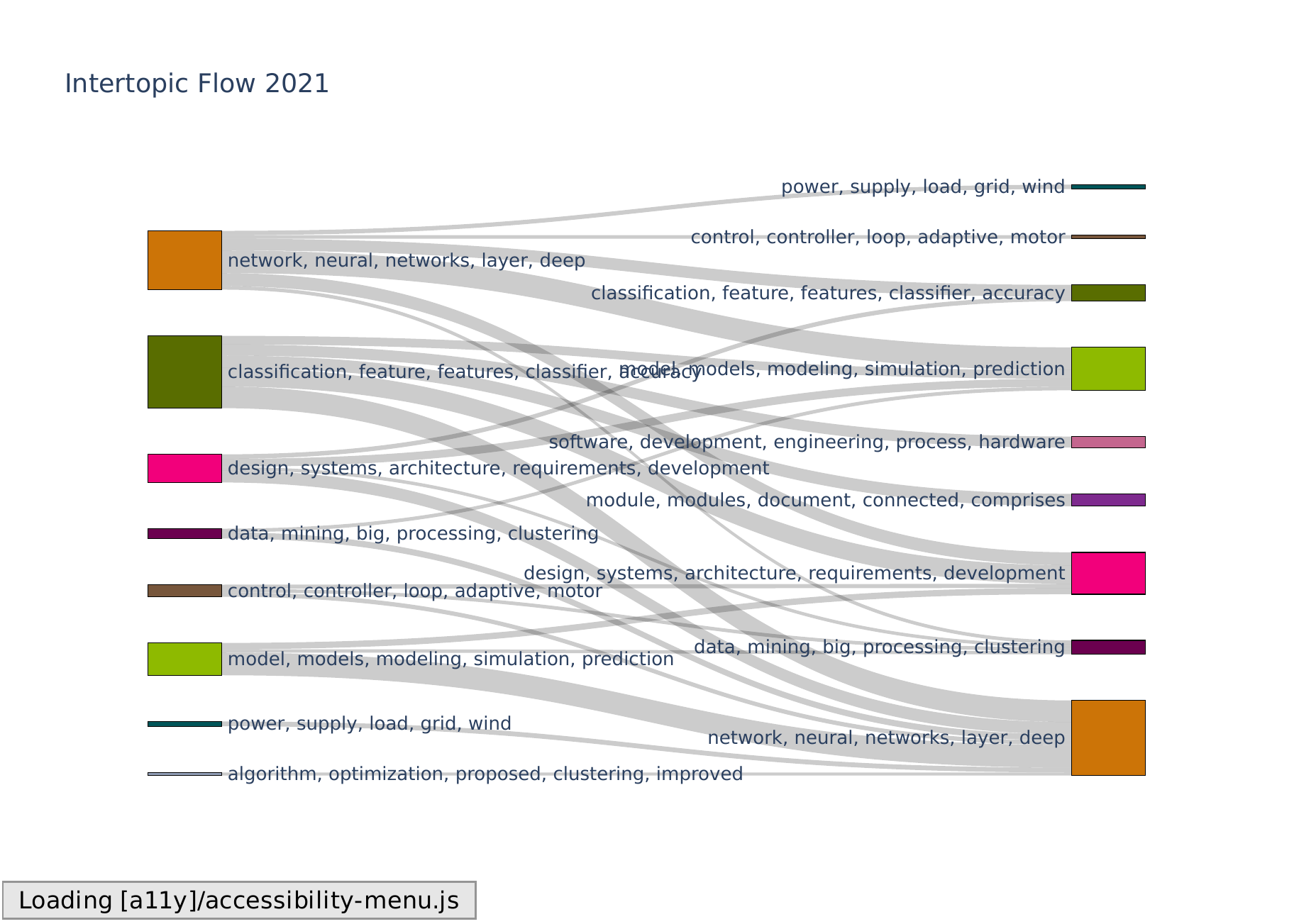}\put (1,59) {Computer Science 2021}\end{overpic}
\caption{Intertopic flows in computer science $C_{\text{CS}}$, 2021.}\label{fig:flowcs2021}
\end{figure}

\begin{figure}
\centering
\begin{overpic}[width=0.9\columnwidth,trim={2cm 1.95cm 2cm 2.4cm},clip]{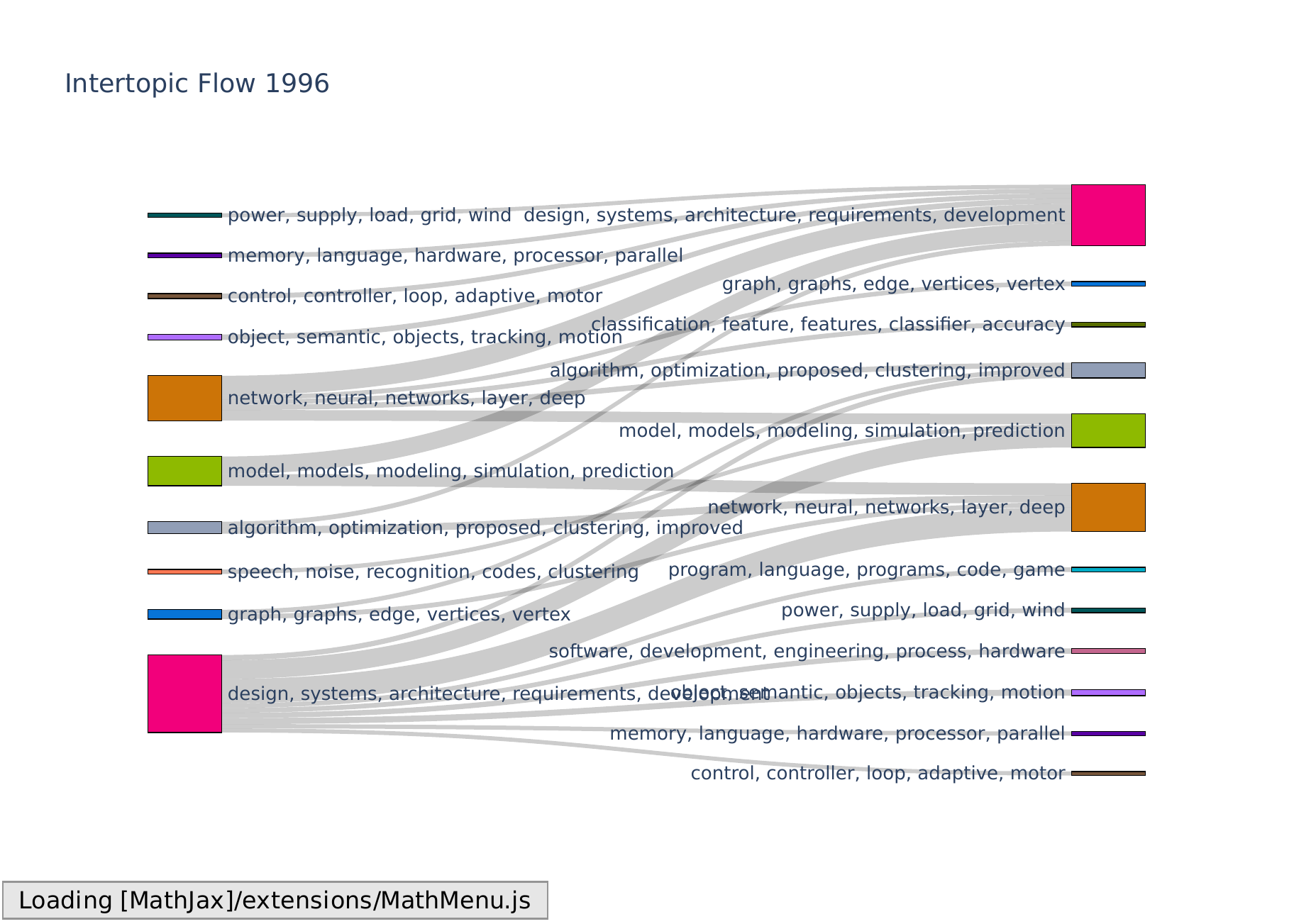}\put (1,59) {Computer Science 1996}\end{overpic}
\caption{Intertopic flows in computer science $C_{\text{CS}}$, 1996.}\label{fig:flowcs1996}
\end{figure}

\begin{figure}
\centering
\begin{overpic}[width=0.9\columnwidth,trim={2cm 1.95cm 2cm 2.4cm},clip]{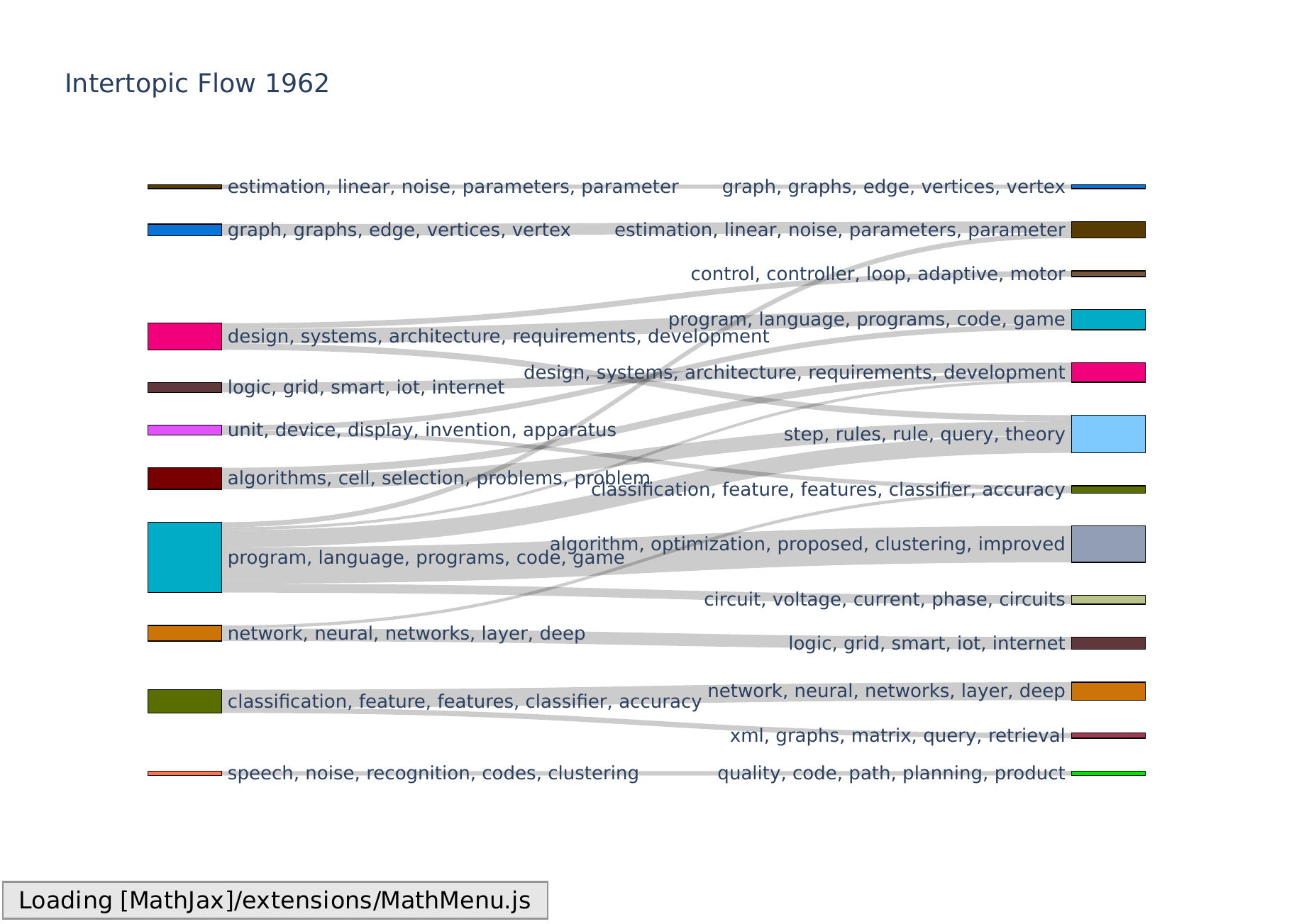}\put (1,61) {Computer Science 1962}\end{overpic}
\caption{Intertopic flows in computer science $C_{\text{CS}}$, 1962.}\label{fig:flowcs1962}
\end{figure}

In our final analysis, we compute intertopic flows as explained in \cref{section:intertopicflows} and visualize them for different years using Sankey flow diagrams. In all our visualizations, source topics are displayed on the left and target topics on the right. The size of an edge connecting a source with a target topic indicates the amount of expertise on the source topic that flows to the target topic.
To obtain a better overview, we depict only the strongest 25 intertopic flows. We exclude intratopic flows as they are responsible for the major part of the flow and would obscure intertopic flows.

In \cref{fig:flowcs2021}, we depict the results for $C_{\text{CS}}$ in 2021. Clearly, neural networks and classification are source topics with strong outgoing flow to a variety of different research topics. Similarly, the neural networks topic is also frequently a target topic. Some of the target topics of neural networks include \emph{simulation, prediction} and \emph{classification}, but also ``practical'' topics such as \emph{power, supply, load, grid, wind}. This may indicate that in 2021, neural networks are already applied in practical scenarios, such as the prediction of wind energy production. \Cref{fig:flowcs1996} shows a substantially different view on $C_{\text{CS}}$ in 1996. We find that the large source topic \emph{classification} is missing in 1996 and all source topics are differently pronounced. The \emph{algorithm} topic was fourth largest target topic and contributed also largely to neural networks.
As a third example for $C_{\text{CS}}$, \cref{fig:flowcs1962} depicts intertopic flows for 1962. At that point in time neural networks were not yet of as much importance as compared to the contemporary status. Overall, intertopic flows occurred between more traditional and basic computer science topics, such as from programming languages (\emph{program, language, programs, code, game}) to algorithms (\emph{algorithm, optimization, proposed, clustering, improved}).

\begin{figure}
\centering
\begin{overpic}[width=0.9\columnwidth,trim={2cm 1.95cm 2cm 2.4cm},clip]{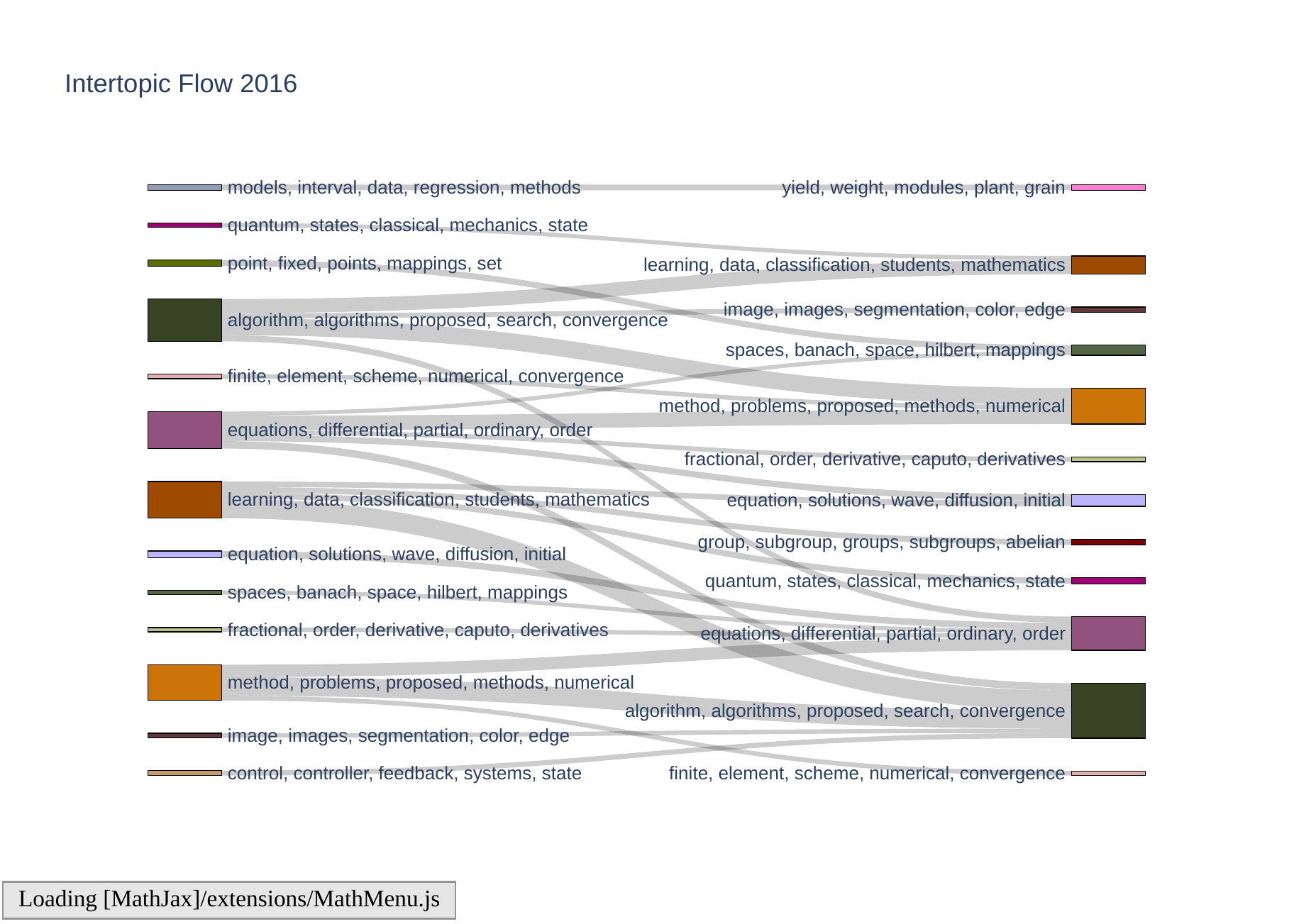}\put (1,61) {Math 2016}\end{overpic}
\caption{Intertopic flows in math $C_{\text{MATH}}$, 2016.}\label{fig:flowmath2016}
\end{figure}
We applied the same analysis of intertopic flows to the \tfns{} resulting from the math data set $C_{\text{MATH}}$. As an example the results for the year 2016 are depicted in \cref{fig:flowmath2016}. In this, we can observe, that the algorithm topic (\emph{algorithm, algorithms, proposed, search, convergence}) is a source of strong intertopic flow, which may be related to the mathematical investigation of algorithms, e.g., concerning convergence properties. Some topics, such as \emph{method, problems, proposed, methods, numerical} are ambiguous. However, taking the incoming flow into account, i.e., \emph{equations, differential, partial, ordinary, order} and \emph{algorithm, algorithms, proposed, search, convergence}, it is revealed that the methods topic might be related to the numerical treatment of differential equations. 
\begin{figure}
\centering
\begin{overpic}[width=0.9\columnwidth,trim={2cm 1.95cm 2cm 2.4cm},clip]{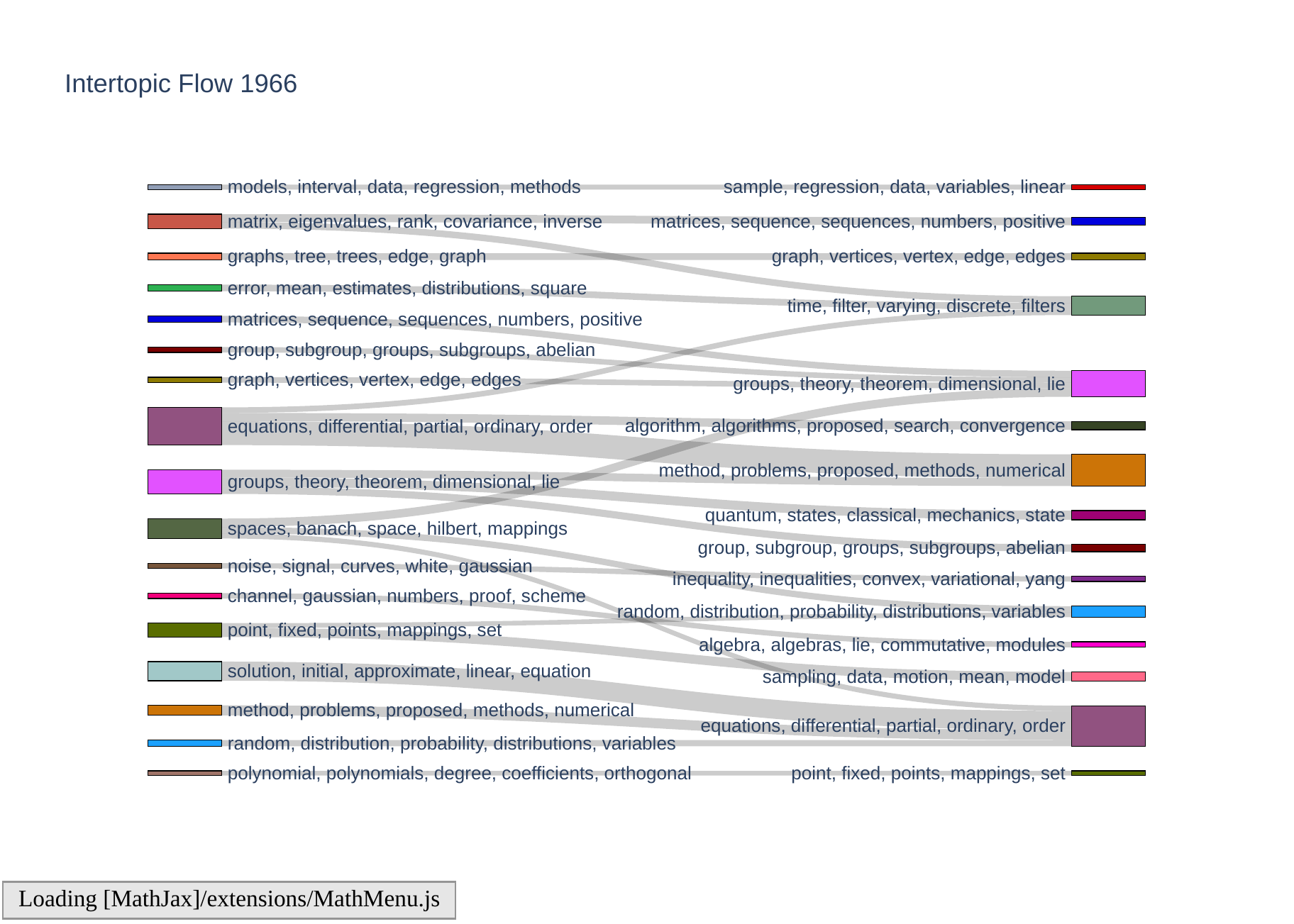}\put (1,61) {Math 1966}\end{overpic}
\caption{Intertopic flows in math $C_{\text{MATH}}$, 1966.}\label{fig:flowmath1966}
\end{figure}
As \cref{fig:flowmath1966} shows, in 1966, i.e., 50 years earlier, there is a considerable difference in intertopic flow compared to 2016.
For example, we find that topics which generate much flow to other topics are \emph{groups, theory, theorem, dimensional, lie} and, again, \emph{equations, differential, partial, ordinary, order}. We find that there is a strong flow from group theory (\emph{groups, theory, theorem, dimensional, lie}) to a topic that we identify as mathematical physics (\emph{quantum, states, classical, mechanics, state}). Thus, our method identifies an influence, which is confirmed by the scientific literature. Moreover, we find further intertopic flows that are supported by literature, e.g., from \emph{graph, vertices, vertex, edge, edges} to \emph{groups, theory, theorem, dimensional, lie}, which we attribute to research about modern algebraic graph theory, or from \emph{random, distribution, probability, distributions, variables} to \emph{equations, differential, partial, ordinary, order}, which might be a consequence of the introduction of probabilistic methods for the solution of differential equations.

In our case study, we computed intertopic flow visualizations for $C_{\text{MATH}}$ and $C_{\text{CS}}$ for more than 60 years of research. Hence, a thorough investigation into all the computed flows requires a separate study and is out of scope of this methodical introduction work into \tfn s. However, we claim that the depicted examples provide enough evidence that the proposed method of \tfns{} is suitable for capturing, visualizing and investigating intertopic flows.


\section{Conclusion \& Outlook}
In this work we investigated the exchange of topic specific
expertise in large scientific collaboration networks.  For this, we
introduced \tfn{}, i.e., an edge weighted multi-graph,
that encodes topical collaborations over time. The edge weights in this \tfns{} result
from the topical collaborations, i.e., research papers.
\tfn s not only allow for an investigation of topic flows, but
their structure also enables analyses with standard methods from graph
theory and social network analysis. Our method requires solely the availability of co-authorship information and  paper abstracts, i.e., data sources that are commonly easier to obtain compared to, e.g., citation data.

To demonstrate the overall applicability
of our approach, we conducted experiments on two large research
corpora from the domains computer science and mathematics. Both corpora were
extracted from the Semantic Scholar Open Research Corpus and span over more than
sixty years of research. We applied several graph based analysis methods to the resulting \tfns s, 
such as PageRank, $k$-cores and community detection. These analyses provide evidence that the introduced graph structure is capable of capturing novel aspects of (topical) collaboration, which were unattainable by the state of the art. A particular unique feature of our method is the ability to uncover collaboration-based intertopic flows.
Most interestingly, and a potential starting point for a broad intertopic study, are the strong differences in flow over time and their respective topics.

%
%
%
For future work, we identified several lines of research. 
First, we restricted some of our investigations to main topics, and therefore main
topics edges. An inclusion of all edges may result in a more detailed
view on intertopic flow. However, the number of the therewith required computations grows in the size of the graph per additional topic.
Second, our
investigations were so far only targeted at capturing and quantifying topic flow. Yet, it could be beneficial to study the causal effects of flow within the collaboration network. Using this, one could investigate influences between research topics over time, e.g., neural networks on computer vision.
Third, the introduced characterization of inter- and intratopic flow does not account for the absolute difference of topical expertise in the \tfns{}. By incorporating this as a weight a more complete picture of the global intertopic flow might emerge.
Once again, this is associated with an increase in computation costs.
Finally, we may note that, although our networks only require co-authorship information, they can be extended to include citation information in a natural way. This in turn would allow for the analysis of the transfer of topical expertise, through collaboration and citation at the same time.

\printbibliography

\newpage

\appendix
\renewcommand{\thesection}{\Alph{section}.\arabic{section}}
\section{Additional Plots and Topic Descriptions}


\begin{table}
  \centering
\caption{Found topics for the math corpus $C_{\text{MATH}}$. For each topic, top five terms are given in order of relevance.}
\begingroup
\setlength{\tabcolsep}{1.2pt}
\renewcommand{\arraystretch}{0.9}
\begin{tabular}{l l l l l l} 
\toprule
\textbf{Topic 1} & \textbf{Topic 2} & \textbf{Topic 3} & \textbf{Topic 4} & \textbf{Topic 5} & \textbf{Topic 6}\\
\toprule
sample & boundary & estimator & space & entropy & yield\\
regression & wave & estimators & detection & bound & weight\\
data & stress & likelihood & product & asymptotic & modules\\
variables & elastic & estimation & tensor & order & plant\\
linear & stochastic & maximum & text & heat & grain\\
\vspace{-2mm}&&&&\\
\toprule
\textbf{Topic 7} & \textbf{Topic 8} & \textbf{Topic 9} & \textbf{Topic 10} & \textbf{Topic 11} & \textbf{Topic 12}\\
\toprule
integral & parameter & function & existence & matrices & polynomial\\
operators & parameters & density & conditions & sequence & polynomials\\
type & distribution & kernel & sufficient & sequences & degree\\
inequalities & hopf & lambda & periodic & numbers & coefficients\\
integrals & bifurcation & constant & condition & positive & orthogonal\\
\vspace{-2mm}&&&&\\
\toprule
\textbf{Topic 13} & \textbf{Topic 14} & \textbf{Topic 15} & \textbf{Topic 16} & \textbf{Topic 17} & \textbf{Topic 18}\\
\toprule
equation & stability & test & group & fuzzy & time\\
solutions & delay & tests & subgroup & membership & filter\\
wave & lyapunov & testing & groups & controller & varying\\
diffusion & delays & statistic & subgroups & interval & discrete\\
initial & functional & hypothesis & abelian & clustering & filters\\
\vspace{-2mm}&&&&\\
\toprule
\textbf{Topic 19} & \textbf{Topic 20} & \textbf{Topic 21} & \textbf{Topic 22} & \textbf{Topic 23} & \textbf{Topic 24}\\
\toprule
graphs & network & graph & channel & let & chapter\\
tree & optimization & vertices & gaussian & denote & series\\
trees & networks & vertex & numbers & ring & section\\
edge & neural & edge & proof & prime & discusses\\
graph & methods & edges & scheme & integer & theory\\
\vspace{-2mm}&&&&\\
\toprule
\textbf{Topic 25} & \textbf{Topic 26} & \textbf{Topic 27} & \textbf{Topic 28} & \textbf{Topic 29} & \textbf{Topic 30}\\
\toprule
soil & finite & watermark & solution & learning & spaces\\
correlation & element & watermarking & initial & data & banach\\
water & scheme & image & approximate & classification & space\\
logic & numerical & embedding & linear & students & hilbert\\
cell & convergence & attacks & equation & mathematics & mappings\\
\vspace{-2mm}&&&&\\
\toprule
\textbf{Topic 31} & \textbf{Topic 32} & \textbf{Topic 33} & \textbf{Topic 34} & \textbf{Topic 35} & \textbf{Topic 36}\\
\toprule
fractional & formula & image & manifold & operator & groups\\
order & number & images & curvature & bounded & theory\\
derivative & text & segmentation & boundary & operators & theorem\\
caputo & numbers & color & distance & norm & dimensional\\
derivatives & formulas & edge & surfaces & convex & lie\\
\vspace{-2mm}&&&&\\
\toprule
\textbf{Topic 37} & \textbf{Topic 38} & \textbf{Topic 39} & \textbf{Topic 40} & \textbf{Topic 41} & \textbf{Topic 42}\\
\toprule
wavelet & flow & optimal & nonlinear & models & method\\
transform & fluid & domain & solutions & interval & problems\\
image & flows & state & semigroup & data & proposed\\
fourier & velocity & regularization & property & regression & methods\\
coefficients & pressure & designs & bifurcation & methods & numerical\\
\vspace{-2mm}&&&&\\
\toprule
\textbf{Topic 43} & \textbf{Topic 44} & \textbf{Topic 45} & \textbf{Topic 46} & \textbf{Topic 47} & \textbf{Topic 48}\\
\toprule
inequality & structure & error & frequency & codes & algebra\\
inequalities & particle & mean & signal & code & algebras\\
convex & positive & estimates & phase & decoding & lie\\
variational & optimization & distributions & signals & binary & commutative\\
yang & equilibrium & square & power & ldpc & modules\\
\vspace{-2mm}&&&&\\
\toprule
\textbf{Topic 49} & \textbf{Topic 50} & \textbf{Topic 51} & \textbf{Topic 52} & \textbf{Topic 53} & \textbf{Topic 54}\\
\toprule
matrix & sets & random & model & control & noise\\
eigenvalues & logic & distribution & linear & controller & signal\\
rank & fuzzy & probability & process & feedback & curves\\
covariance & decision & distributions & stochastic & systems & white\\
inverse & set & variables & markov & state & gaussian\\
\vspace{-2mm}&&&&\\
\toprule
\textbf{Topic 55} & \textbf{Topic 56} & \textbf{Topic 57} & \textbf{Topic 58} & \textbf{Topic 59} & \textbf{Topic 60}\\
\toprule
functions & problem & sampling & operators & equations & quantum\\
analytic & sequence & data & topological & differential & states\\
symmetric & solving & motion & manifolds & partial & classical\\
real & programming & mean & compact & ordinary & mechanics\\
class & solve & model & metric & order & state\\
\vspace{-2mm}&&&&\\
\toprule
\textbf{Topic 61} & \textbf{Topic 62} & \textbf{Topic 63} & \textbf{Topic 64}\\
\toprule
complexity & algorithm & systems & point\\
operators & algorithms & chaotic & fixed\\
computational & proposed & dynamical & points\\
algorithms & search & synchronization & mappings\\
problems & convergence & chaos & set\\
\vspace{-2mm}&&&&\\
\end{tabular}
\endgroup
\label{table:mathtopics}
\end{table}

\begin{table}[h!]
\centering
\caption{Found topics for the Computer Science corpus $C_{\text{CS}}$. For each topic, top five terms are given in order of relevance.}
\begingroup
\setlength{\tabcolsep}{1.2pt}
\renewcommand{\arraystretch}{0.9}
\begin{tabular}{l l l l l l} 
\toprule
\textbf{Topic 1} & \textbf{Topic 2} & \textbf{Topic 3} & \textbf{Topic 4} & \textbf{Topic 5} & \textbf{Topic 6}\\
\toprule
face & teaching & net & program & social & sensor\\
recognition & course & petri & language & network & nodes\\
facial & english & nets & programs & users & node\\
faces & students & state & code & packet & sensors\\
expression & teachers & object & game & media & wireless\\
\vspace{-2mm}&&&&\\
\toprule
\textbf{Topic 7} & \textbf{Topic 8} & \textbf{Topic 9} & \textbf{Topic 10} & \textbf{Topic 11} & \textbf{Topic 12}\\
\toprule
data & estimation & power & search & fuzzy & user\\
mining & linear & supply & query & decision & users\\
big & noise & load & engine & membership & interface\\
processing & parameters & grid & engines & controller & authentication\\
clustering & parameter & wind & quantum & sets & device\\
\vspace{-2mm}&&&&\\
\toprule
\textbf{Topic 13} & \textbf{Topic 14} & \textbf{Topic 15} & \textbf{Topic 16} & \textbf{Topic 17} & \textbf{Topic 18}\\
\toprule
vehicle & routing & management & algorithms & graph & antenna\\
vehicles & protocol & business & cell & graphs & controller\\
driving & protocols & chapter & selection & edge & ghz\\
road & nodes & information & problems & vertices & antennas\\
driver & hoc & mobile & problem & vertex & array\\
\vspace{-2mm}&&&&\\
\toprule
\textbf{Topic 19} & \textbf{Topic 20} & \textbf{Topic 21} & \textbf{Topic 22} & \textbf{Topic 23} & \textbf{Topic 24}\\
\toprule
speech & students & security & design & model & image\\
noise & knowledge & privacy & systems & models & images\\
recognition & student & attacks & architecture & modeling & color\\
codes & research & authentication & requirements & simulation & segmentation\\
clustering & skills & secure & development & prediction & processing\\
\vspace{-2mm}&&&&\\
\toprule
\textbf{Topic 25} & \textbf{Topic 26} & \textbf{Topic 27} & \textbf{Topic 28} & \textbf{Topic 29} & \textbf{Topic 30}\\
\toprule
memory & robot & xml & channel & communication & cloud\\
language & mobile & graphs & channels & wireless & computing\\
hardware & robots & matrix & scheme & terminal & storage\\
processor & motion & query & mimo & mobile & encryption\\
parallel & human & retrieval & fading & information & information\\
\vspace{-2mm}&&&&\\
\toprule
\textbf{Topic 31} & \textbf{Topic 32} & \textbf{Topic 33} & \textbf{Topic 34} & \textbf{Topic 35} & \textbf{Topic 36}\\
\toprule
circuit & task & logic & information & training & unit\\
voltage & genetic & grid & online & human & device\\
current & tasks & smart & content & methods & display\\
phase & policy & iot & education & dataset & invention\\
circuits & policies & internet & technology & samples & apparatus\\
\vspace{-2mm}&&&&\\
\toprule
\textbf{Topic 37} & \textbf{Topic 38} & \textbf{Topic 39} & \textbf{Topic 40} & \textbf{Topic 41} & \textbf{Topic 42}\\
\toprule
learning & step & software & energy & algorithm & network\\
machine & rules & development & consumption & optimization & neural\\
deep & rule & engineering & wireless & proposed & networks\\
virtual & query & process & efficiency & clustering & layer\\
learners & theory & hardware & battery & improved & deep\\
\vspace{-2mm}&&&&\\
\toprule
\textbf{Topic 43} & \textbf{Topic 44} & \textbf{Topic 45} & \textbf{Topic 46} & \textbf{Topic 47} & \textbf{Topic 48}\\
\toprule
module & service & water & traffic & quality & scheduling\\
modules & services & forecasting & road & code & resource\\
document & qos & sar & images & path & allocation\\
connected & quality & equations & segmentation & planning & problem\\
comprises & business & radar & flow & product & resources\\
\vspace{-2mm}&&&&\\
\toprule
\textbf{Topic 49} & \textbf{Topic 50} & \textbf{Topic 51} & \textbf{Topic 52} & \textbf{Topic 53} & \textbf{Topic 54}\\
\toprule
book & method & input & frequency & library & control\\
books & filter & output & signal & digital & controller\\
download & nonlinear & mode & rate & libraries & loop\\
reading & numerical & device & bit & quantum & adaptive\\
like & finite & converter & ofdm & resources & motor\\
\vspace{-2mm}&&&&\\
\toprule
\textbf{Topic 55} & \textbf{Topic 56} & \textbf{Topic 57} & \textbf{Topic 58} & \textbf{Topic 59} & \textbf{Topic 60}\\
\toprule
health & web & fault & agent & detection & video\\
server & pages & parallel & agents & intrusion & coding\\
database & page & faults & multi & signal & frame\\
care & decision & diagnosis & measurement & flow & signal\\
monitoring & services & tolerant & decision & layer & audio\\
\vspace{-2mm}&&&&\\
\toprule
\textbf{Topic 61} & \textbf{Topic 62} & \textbf{Topic 63} & \textbf{Topic 64}\\
\toprule
fusion & test & object & classification\\
cache & testing & semantic & feature\\
multi & optimization & objects & features\\
fused & problem & tracking & classifier\\
local & problems & motion & accuracy\\
\vspace{-2mm}&&&&\\
\end{tabular}
\endgroup
\label{table:cstopics}
\end{table}

\begin{figure}
\includegraphics[width=\columnwidth,trim={0.9cm 1cm 5.5cm 2cm},clip]{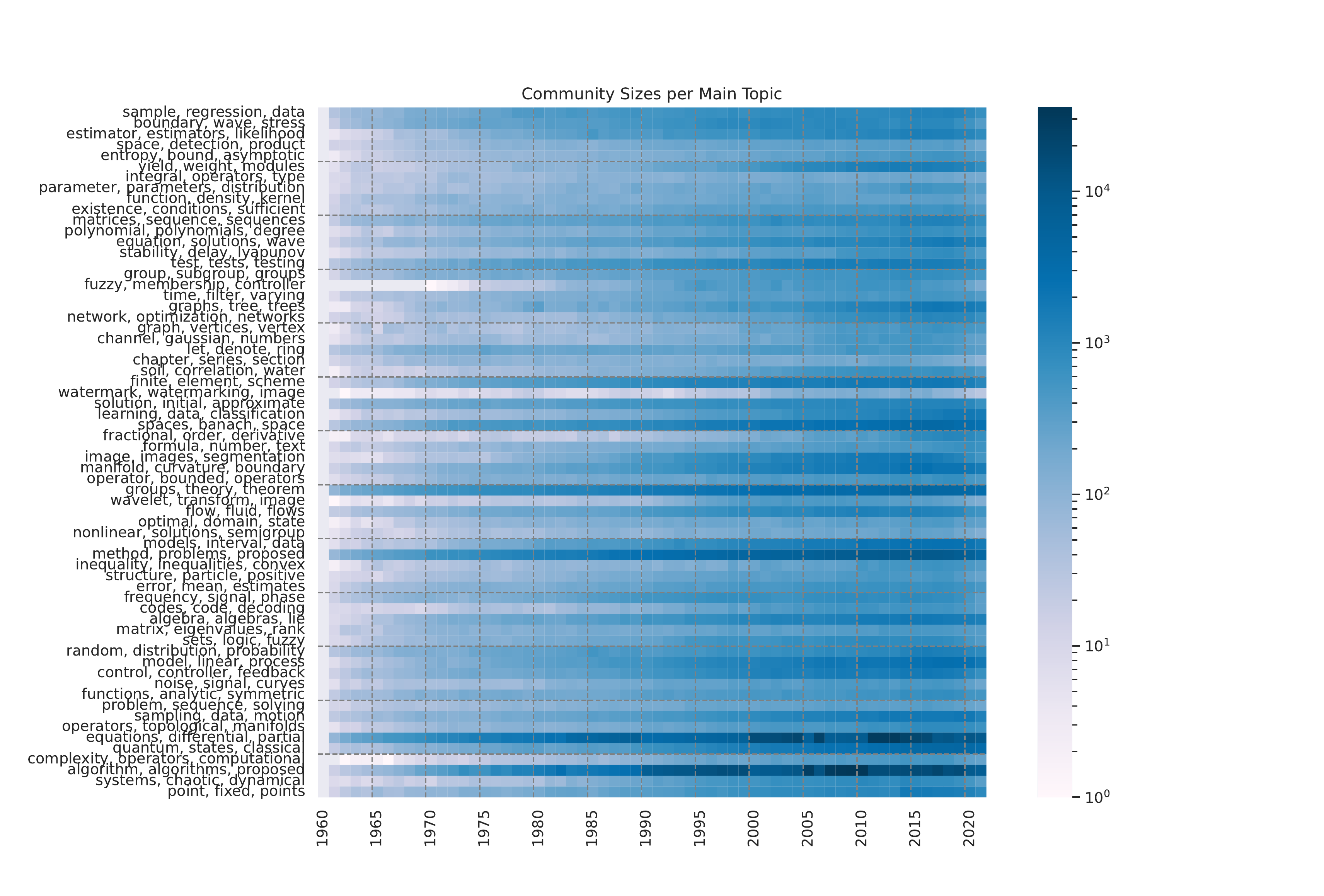}
\caption{Community sizes for math data set.}\label{fig:communitiesmath}
\end{figure}

\begin{figure}
\includegraphics[width=\columnwidth,trim={0.9cm 1cm 5.5cm 2cm},clip]{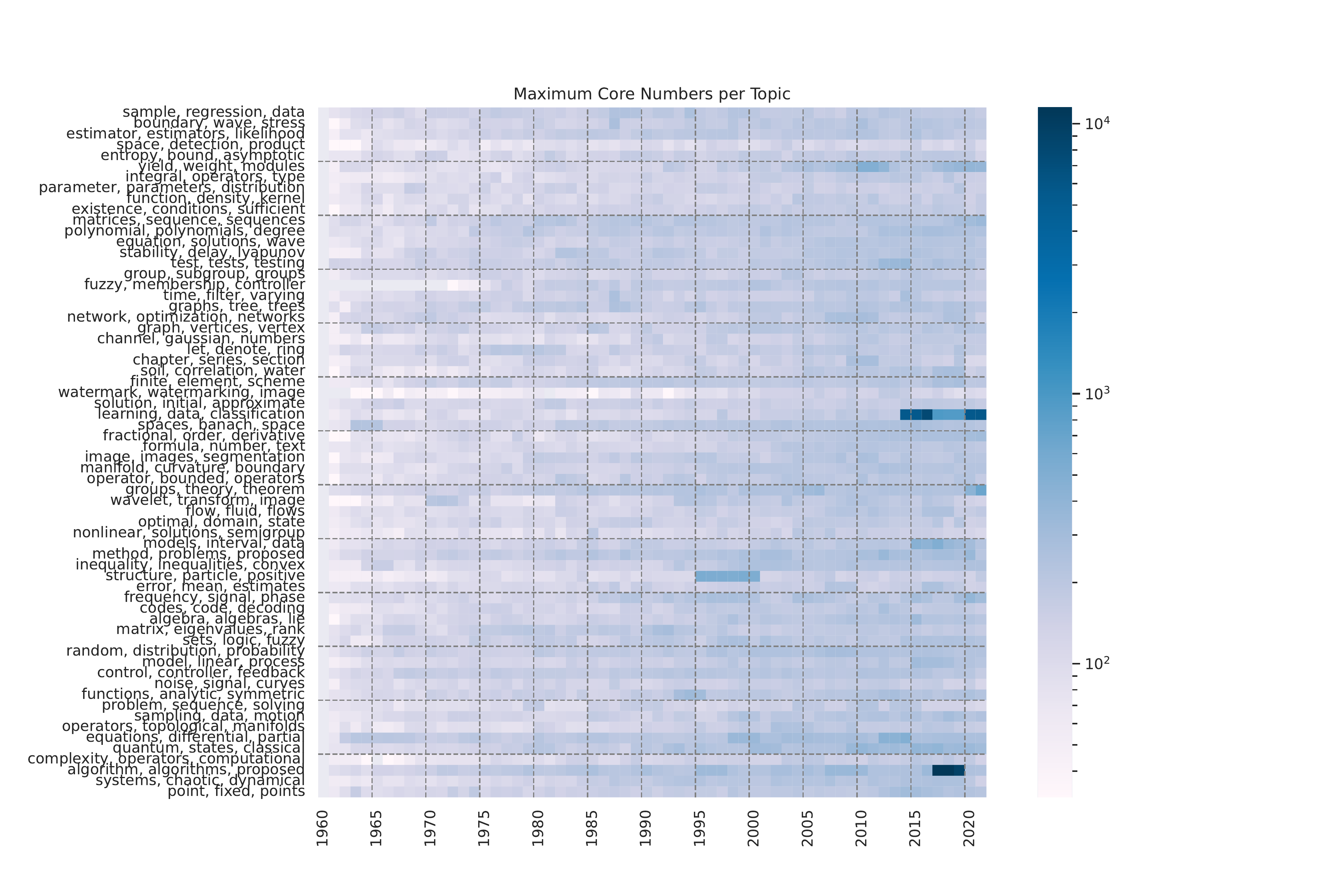}
\caption{Maximum core numbers for math data set.}\label{fig:coresmath}
\end{figure}


\end{document}